\documentclass[aps,print,prb,twocolumn,superscriptaddress]{revtex4-2}

\usepackage[hidelinks]{hyperref}
\hypersetup{
	colorlinks,
	linkcolor={red},
	citecolor={blue},
	urlcolor={blue}
}
\usepackage{pifont}
\usepackage{balance}
\usepackage{amssymb}
\usepackage{suffix}
\usepackage{mathtools}
\usepackage[utf8]{inputenc}
\usepackage{booktabs}
\usepackage{cases}
\usepackage[multiple]{footmisc}
\usepackage{dcolumn}
\usepackage{color,soul}\usepackage{tabularx}
\usepackage{rotating}
\usepackage{perpage}
\usepackage{siunitx}
\usepackage{xcolor}
\usepackage{soul}
\usepackage{amsmath}
\usepackage{tikz}
\usepackage[T1]{fontenc}
\usepackage{etoolbox}
\usepackage{graphics}
\usepackage{siunitx}
\usepackage{float}	
\usepackage{collref}
\usepackage{multirow}
\usepackage{mathtools}
\usepackage{bm}
\usepackage{url}

\usepackage{tikz}
\usepackage{tikz-3dplot}
\usepackage{accents}

\usepackage{pgfplots}
\pgfplotsset{compat=1.18}
\makeatletter
\newcommand{\doublewidetilde}[1]{{%
		\mathpalette\double@widetilde{#1}%
}}
\newcommand{\double@widetilde}[2]{%
	\sbox\z@{$\m@th#1\widetilde{#2}$}%
	\ht\z@=.9\ht\z@
	\widetilde{\box\z@}%
}
\makeatother
\usepackage{hyperref}  

\newcommand{\cmark}{\textcolor{green!60!black}{\scalebox{1.7}{\ding{51}}}}
\newcommand{\xmark}{\textcolor{red!70!black}{\scalebox{1.7}{\ding{55}}}}

\usepackage{etoolbox,lipsum}

\newcommand{\red}[1]{\textcolor{black}{#1}}
\newcommand{\redd}[1]{\textcolor{black}{#1}}

\newcommand{\rd}[1]{\textcolor{black}{#1}}

\begin{document} 
	
	\title{Efficient two-color Floquet control of the RKKY interaction in altermagnets}
	
	\author{Mohsen Yarmohammadi}
	\email{mohsen.yarmohammadi@georgetown.edu}
	\address{Department of Physics, Georgetown University, Washington DC 20057, USA}
	\author{Pei-Hao Fu}
	\address{Department of Physics and Astronomy, University of Florence, I-50019 Sesto Fiorentino, Italy}
	\author{James K. Freericks}
	\address{Department of Physics, Georgetown University, Washington DC 20057, USA}
	
	\date{\today}
	
	\begin{abstract}
		Magnetic impurities in real materials can mask the intrinsic spin-dependent properties of hosts. They interact indirectly through the Ruderman–Kittel–Kasuya–Yosida~(RKKY) mechanism, which limits the use of isolated impurity spins in applications such as qubits and spintronics. By suppressing the RKKY interaction, an effective control scheme should therefore enable access to the host’s unperturbed behavior while simultaneously isolating the impurity spins for functional use. Although beyond static approaches, single-color laser driving can suppress the RKKY interaction, it typically requires strong fields that may be impractical or destabilizing. To overcome these limitations, we show that two-color laser driving provides efficient and tunable control over all components of the RKKY interaction using two weak laser fields. Focusing on two-dimensional Rashba altermagnets, we show that interference between one- and two-photon processes produces altermagnet-specific Floquet corrections. These include additional AC Stark shifts, magnetizations, spin–orbit renormalization, and emergent in-plane Zeeman fields, which are absent under single-color driving and in non-altermagnetic systems. Notably, two-color driving induces a finite \(z\)-component of the Dzyaloshinskii--Moriya (DM) interaction, stabilizing in-plane chiral magnetism and related textures in Rashba altermagnets. These effects enable tunable, near-complete on--off switching of the Heisenberg, Ising, and DM interactions through a Lifshitz-like modulation of the Fermi surface. We also show that the tuning process is highly sensitive to the chirality of both beams. We further map out phase diagrams for ferromagnetic and antiferromagnetic alignment of impurities with clockwise and counterclockwise canting as a function of Rashba coupling and altermagnetic order. Finally, we discuss candidate material platforms and experimental feasibility.
	\end{abstract}
	
	\maketitle
	{\allowdisplaybreaks
		\section{Introduction}
		
		Altermagnets are a recently identified class of magnetic materials~\cite{PhysRevX.12.040501,PhysRevX.12.031042,PhysRevX.12.040002,Krempaský2024} whose electronic bands exhibit momentum-dependent spin splitting despite vanishing net magnetization. Unlike ferromagnets, where spin splitting arises from uniform magnetization, or conventional antiferromagnets, where combined symmetries enforce spin degeneracy in momentum space, altermagnets host spin-polarized bands with spin polarization that changes sign across the Brillouin zone~\cite{PhysRevX.12.040501,PhysRevX.12.031042,PhysRevX.12.040002,Krempaský2024,doi:10.1126/sciadv.aaz8809,PhysRevB.99.184432,Naka2019,doi:10.7566/JPSJ.88.123702,PhysRevX.12.011028,PhysRevLett.128.197202,PhysRevLett.130.216701}. This distinct symmetry structure gives rise to unconventional transport and optical responses \cite{Jungwirth2024,Bai2024,Jungwirth2024a,Yuri2025Superconducting}, including charge–spin conversion~\cite{Bose2022}, giant and tunneling magnetoresistance~\cite{PhysRevX.12.011028}, anomalous Hall effects~\cite{doi:10.1126/sciadv.adn0479,Feng2022,doi:10.1126/sciadv.aaz8809}, nonlinear transport \cite{nonlinearghorashi}, spin filtering effects \cite{Bai2024, Yan2024, Fu2025All, Herasymchuk2025Electric, Zhu2025Altermagnetoelectric, Kokkeler2025Nonequilibrium, Yang2025Altermagnetic, Yang2025Unconventional, Fang2025Edgetronics, Peng2025Ferroelastic} among other phenomena \cite{Schwartz2025Thermomagnonic, Yoshida2025Quantization, Golub2025Edge, Sunko2025Linear, Maznichenko2024Fragile}, while maintaining magnetic compensation. Altermagnets therefore provide a promising platform for spintronic functionalities without macroscopic magnetization~\cite{PhysRevLett.129.137201,PhysRevLett.134.106801,PhysRevLett.134.106802,PhysRevB.110.235101}.

		Because the defining property of altermagnets lies in the momentum-resolved spin structure of their electronic states~\cite{PhysRevB.102.075112,PhysRevB.103.125114,Feng2022,Reichlova2024,PhysRevB.109.094413}, it is natural to ask how this spin texture influences magnetic interactions mediated by itinerant electrons. Magnetic impurities embedded in metallic hosts provide a transparent platform for addressing this question. In particular, the Ruderman–Kittel–Kasuya–Yosida (RKKY) interaction~\cite{10.1143/PTP.16.45,PhysRev.106.893,PhysRev.96.99,Lee2025,PhysRevB.110.054427} generates an indirect exchange between spatially separated local spins and is highly sensitive to the host band structure, Fermi-surface geometry, and spin-dependent correlations. Analyzing the RKKY interaction, therefore, provides direct insight into how altermagnetic spin textures reshape impurity-mediated magnetic coupling at the nanoscale~\cite{Lee2025,PhysRevB.110.054427,k3xb-8pts,Lee2025,kundu2025rkkyinteractionmediatedspinpolarized,duan2026neelvectorrashbasoc}.
		
		Rashba altermagnets, where Rashba spin–orbit coupling (RSOC) coexists with altermagnetic order, offer a versatile platform for engineering spin-dependent interactions. While the RKKY interaction in these systems is highly anisotropic and tunable~\cite{PhysRevB.110.054427,k3xb-8pts,duan2026neelvectorrashbasoc}, controllable on–off switching remains challenging. Magnetic impurities in real materials inevitably obscure their intrinsic behavior. To accurately probe the host’s fundamental behavior, it is therefore important to minimize or account for their influence while recognizing that the impurities cannot be eliminated. On the other hand, although RKKY-mediated exchange between impurities can stabilize collective magnetic phases, its long-range nature often masks single-impurity physics. Individual magnetic impurities are technologically significant, acting as addressable quantum spins in quantum information science~\cite{Pla2012,Gilbert2023}, as ultrasensitive nanoscale sensors~\cite{Bonizzoni2024,Hache2025,WRACHTRUP2016225}, and as functional elements in spintronic memory devices~\cite{Zhao2025,https://doi.org/10.1002/adma.202202841}. Suppressing RKKY coupling enables effectively decoupled local spins, opening the door to single-spin applications.

        Given this background, an efficient control strategy is essential both to (i) extract pristine information from the altermagnet in realistic experimental setups with dilute magnetic impurities and (ii) decouple individual magnetic impurities for single-spin device applications. In this context, dynamical control via external periodic driving offers a reversible route to tuning indirect exchange beyond static limitations~\cite{xu2025stoichiometrycontrolledstructuralordertunable,PhysRevB.110.054427}. Optical control of the RKKY interaction has been more explored in various nonequilibrium settings~\cite{PhysRevB.111.014440,PhysRevB.107.054439,PhysRevResearch.2.033228,PhysRevB.110.035307,k3xb-8pts,tm58-lbdl,PhysRevB.100.075127}, including Floquet-engineered band renormalization and magnetizing the system. \rd{Recent studies have made significant strides in evaluating the RKKY interaction using Floquet-Keldysh methods, determining electronic steady states in the presence of a reservoir, and demonstrating light-induced noncollinear exchange in conventional Rashba systems~\cite{Asmar_2021,PhysRevB.90.195429,Asmar_2024}. However, under monochromatic driving, achieving such control generally requires strong fields to appreciably modify the Fermi-level states, which are necessary for efficient tuning of the RKKY interaction to either decouple magnetic impurities or reveal the intrinsic host properties.} This requirement, in turn, limits both the efficiency and robustness of the approach. These limitations motivate the exploration of alternative driving protocols that can reshape the electronic structure more efficiently, enabling control over the RKKY interaction and impurity decoupling without requiring excessively strong fields that risk destabilizing the system.\begin{figure}[t]
        	\centering
        	\includegraphics[width=0.95\linewidth]{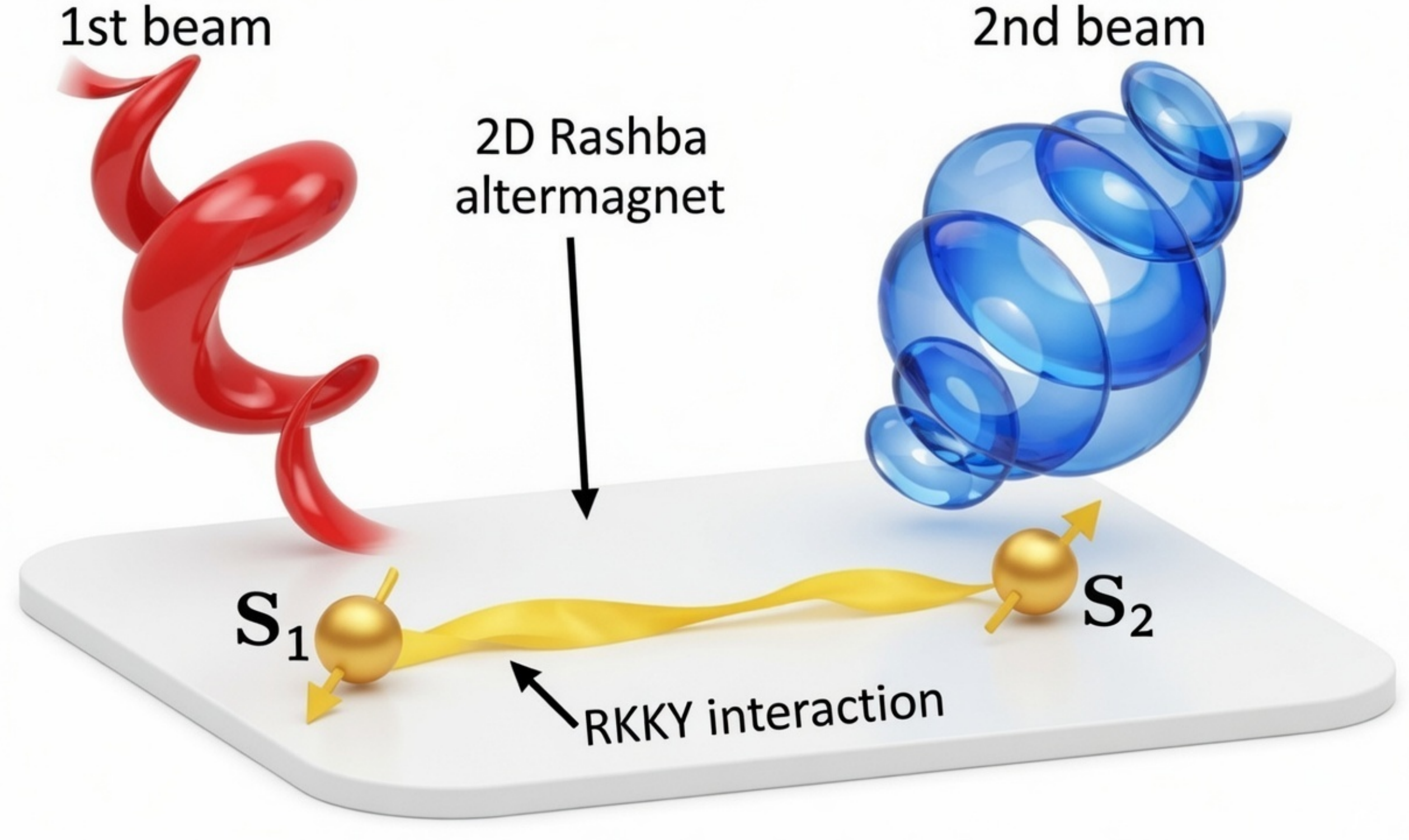}
        	\caption{Schematic illustration of two-color Floquet engineering of the RKKY interaction in a 2D Rashba altermagnet. Two coherent laser beams irradiate the system, with frequencies $\omega_1$ and $\omega_2$, amplitudes $A_0$ and $\mathcal{S}A_0$ (where $\mathcal{S}$ denotes the relative strength of the second-harmonic component), phases $\phi_1$ and $\phi_2$, and chiralities $\eta_1$ and $\eta_2$. Two localized magnetic moments, $\mathbf{S}_1$ and $\mathbf{S}_2$, interact indirectly via the laser-modified itinerant electrons, yielding a tunable RKKY exchange. The relative amplitude ratio and chirality of the beams control the strength, anisotropy, and components of the RKKY coupling.}
        	\label{f1}
        \end{figure}
        
        In this work, we demonstrate that bichromatic driving~\cite{Andiel_1999,Cormier2000,PhysRevLett.94.243901,PhysRevA.73.063823,PhysRevLett.110.233903,Kroh18,Jin15,Jin2014,Jin20014,RAJPOOT2024129241,liu2025inplaneopticallytunablemagnetic,ganguli2025tunabletopologyhallresponse,yarmohammadi2026floquetinducedanisotropicmagnetoresistanceanomalous,yarmohammadi2026giantperpendicularedelsteinpolarization,zhu2026parityselectivespinsplittingcoplanar} provides precisely such a mechanism in altermagnets. In particular, we propose a two-color laser driving protocol in two-dimensional (2D) Rashba altermagnets, using two weak laser beams instead of a single strong beam. The coherent fields, with distinct frequencies, phases, chiralities, and amplitudes, provide enhanced control over Floquet band hybridization and momentum-resolved spin polarization. This enables a much stronger on–off switching process—and in some regimes complete cancellation—of RKKY tensor components compared to monochromatic driving~\cite{k3xb-8pts}. While the mechanism we consider is generally applicable to a broad class of altermagnetic wave patterns—including $d$-, $g$-, $f$-, and $i$-wave forms~\cite{PhysRevX.12.040501,PhysRevX.12.040002,PhysRevX.12.031042,PhysRevB.99.184432,doi:10.7566/JPSJ.88.123702,Reimers2024}—we focus here on $d$-wave altermagnets, which have been predicted to host a variety of unconventional current-driven spin phenomena, such as exotic spin torques, spin--orbit torques, and spin-splitter currents~\cite{PhysRevLett.129.137201,doi:10.1126/sciadv.adn0479}. \rd{The novelty of our approach lies in utilizing bichromatic interference specifically within an altermagnetic host, where the interplay between the two colors and the $d$-wave spin splitting induces novel exchange symmetries that are inaccessible in earlier monochromatic or non-altermagnetic Floquet studies.} 
		
		The remainder of this paper is organized as follows. In Sec.~\ref{s2}, we introduce the model Hamiltonian for 2D Rashba altermagnets and describe the two-color laser driving protocol. In Sec.~\ref{s3}, we present the theoretical framework for evaluating the laser-modified RKKY interaction between magnetic impurities. Section~\ref{s4} contains numerical results illustrating on--off switching of RKKY interaction across a range of model parameters. Section~\ref{s5} provides experimental relevance and candidate materials. Finally, Sec.~\ref{s6} summarizes our findings and briefly outlines prospects for future studies.
		
		\section{Hamiltonian model}\label{s2}
		We consider a 2D Rashba $d$-wave altermagnet described by a minimal tight-binding model capturing kinetic energy, RSOC, and altermagnetic exchange. \red{We employ a quadratic continuum model valid near the $\Gamma$ point.} In equilibrium, these terms generate momentum-dependent spin splitting while preserving zero net magnetization. External periodic driving, as shown in Fig.~\ref{f1}, is incorporated to dynamically reshape the electronic structure and, in turn, the impurity-mediated RKKY interaction later. The equilibrium Hamiltonian is given by $H_{\rm d} = H_{\rm kin} + H_{\rm R} + H_{\rm AM}$, where $H_{\rm kin}$ describes the parabolic band dispersion, $H_{\rm R}$ the RSOC of strength $\lambda_{\rm R}$, and $H_{\rm AM}$ the altermagnetic exchange characterized by $M_{\rm d}$. Thus, the bare conduction electrons in a 2D Rashba altermagnet are described by
		{\small	\begin{align}
				H_\text{d}(\mathbf{k}) &= \varepsilon(k_x^2 + k_y^2)  \sigma_0 
				+ \lambda_{\rm R} (k_x \sigma_y - k_y \sigma_x) 
				+ M_\mathbf{k}^{\rm d} \sigma_z,
		\end{align}}where $\varepsilon = \hbar^2 / 2m_{\rm e}$ is the band mass term. The last term,
		\begin{equation}
			M_\mathbf{k}^{\rm d} = \frac{\hbar^2 M_{\rm d}}{m_{\rm e}} \left[ \frac{(k_x^2 - k_y^2)}{2} \cos(2\theta_M^{\rm d})  + k_x k_y \sin(2\theta_M^{\rm d}) \right],
		\end{equation}represents the momentum-dependent altermagnetic exchange splitting characteristic of $d$-wave altermagnets.  The angle $\theta_M^{\rm d}$ rotates the $d$-wave texture in the Brillouin zone: $\theta_M^{\rm d} = 0$ corresponds to a $d_{x^2-y^2}$-type altermagnetic pattern, while $\theta_M^{\rm d} = \pi/4$  corresponds to a $d_{xy}$-type pattern. The Rashba term induces in-plane spin--momentum locking, with spins oriented perpendicular to momentum ($\textbf{k}\times\boldsymbol{\sigma}$)$_z$, underpinning spin textures and transport. In contrast, the altermagnetic term $M_{\mathbf{k}}^{\rm d}\sigma_z$ vanishes along symmetry lines, and changes sign between opposite momentum sectors, yielding a compensated, noncollinear spin texture in $\textbf{k}$-space with zero net magnetization---the hallmark of altermagnets. Their interplay produces a highly anisotropic momentum-space spin texture.

        Although the system is intrinsically a Rashba altermagnet and already exhibits rich equilibrium properties, external light irradiation offers a versatile and controllable route to further tailor its electronic and spin structure beyond static limitations. In particular, time-periodic driving can dynamically renormalize the band dispersion, spin splitting, and symmetry characteristics without modifying the underlying lattice. Optical excitation can selectively break or reshape discrete symmetries, generate effective spin–orbit fields, and produce nonequilibrium spin textures that do not arise in equilibrium. In this way, light acts as an external tuning parameter for manipulating altermagnetic spin-dependent phenomena in situ, enabling controlled modification of exchange interactions and magnetic anisotropies.

        Although various driving protocols involving linearly polarized fields, or mixed circular and linear polarization, can in principle be considered, we focus here on two circularly polarized lights for physical clarity and minimality. In a 2D Rashba altermagnet, circularly polarized light dynamically breaks time-reversal symmetry and introduces a well-defined chirality, which is essential for activating Floquet-induced exchange channels beyond simple parameter renormalization~\cite{k3xb-8pts,tm58-lbdl}. In contrast, linearly polarized driving preserves an effective time-reversal symmetry over one period and predominantly leads to anisotropic band renormalization without generating qualitatively new interaction terms. Mixed circular–linear schemes produce intermediate effects that can be understood as subsets of the circular–circular case. Thus, two circularly polarized fields constitute the minimal optical configuration that captures the full symmetry-allowed Floquet control of a Rashba altermagnet.
        
        Accordingly, the system is driven by two circularly polarized laser fields, incorporated via the Peierls substitution $\mathbf{k}\rightarrow\mathbf{k}+e\mathbf{A}(t)/\hbar$. The coupling to the electromagnetic field enters through the vector potential $\mathbf{A}(t) = \bigl(A_x(t), A_y(t)\bigr)$ with
		\begin{subequations}
		\redd{\begin{align}
				A_x(t) &= A_0\!\left[\cos(\omega t) +  \mathcal{S}\cos\!\left(n\omega t + \phi\right)\right], \\
				A_y(t) &= A_0\!\left[\eta_1 \sin(\omega t) + \eta_2 \mathcal{S}\sin\!\left(n\omega t + \phi\right)\right].
			\end{align}}
		\end{subequations}In this formulation, the dimensionless ratio $\mathcal{S}$ parametrizes the relative strength of the second harmonic component. The factors $\eta_{1,2}$ encode the chirality of the two laser fields, while $n$ and $\phi = \phi_2 - \phi_1$ denote the frequency ratio and relative phase offset between the driving fields, respectively. The amplitude of the vector potential is further defined as
		\(
		A_0 = E_0 / \omega
		\),
		where \(E_0\) and \(\omega\) denote the electric-field amplitude and frequency of the first laser beam, respectively. In equilibrium, the characteristic momentum-dependent spin splitting of altermagnets is protected by the combined $C_4 \mathcal{T}$ symmetry. Optical driving modifies this symmetry structure. While linearly polarized light reduces rotational symmetry, circularly polarized light explicitly breaks $C_4 \mathcal{T}$ due to its definite chirality.
        
        Although the formalism developed here applies in principle to general bichromatic driving schemes of the form $\omega$--$n\omega$, we focus on the $\omega$--$2\omega$ configuration for the following concrete physical and experimental reasons. The second-harmonic component represents the simplest nonlinear extension beyond single-frequency driving and enables strong interference between single- and two-photon absorption pathways at moderate field strengths. This allows efficient symmetry control and Floquet-induced band mixing without requiring intense electromagnetic fields. \red{Moreover, $\omega$--$2\omega$ driving couples naturally to even-parity and quadrupolar features of the band structure, making it particularly compatible with $d$-wave altermagnetic order. Experimentally, phase-coherent $\omega$--$2\omega$ laser setups are well established~\cite{Andiel_1999,Cormier2000,PhysRevLett.94.243901,PhysRevA.73.063823,PhysRevLett.110.233903}, offering precise control over relative phase, polarization, and intensity.} 

        Since the Hamiltonian is periodic in time, $H_{\rm d}(\textbf{k}, t + 2\pi/\omega) = H_{\rm d}(\textbf{k}, t)$, the driven system can be analyzed using the high-frequency Floquet–Magnus expansion~\cite{GRIFONI1998229,PLATERO20041}, which maps the time-dependent problem onto an effective time-independent eigenvalue equation in an extended Hilbert space. Although periodic driving can, in principle, induce heating, we consider the off-resonant regime with weak driving amplitude, where the system is expected to enter a long-lived prethermal state that is accurately described by the effective Floquet Hamiltonian. In this regime, a controlled analytical treatment becomes possible. Accordingly, we set the driving frequency to \red{$\hbar \omega \geq 1\,\mathrm{eV}$}, exceeding the relevant low-energy electronic bandwidth. As a result, Floquet sidebands remain well separated, resonant crossings are suppressed, and photon-assisted interband transitions are negligible. This separation of energy scales allows averaging over the rapid micromotion induced by the periodic drive, yielding an effective static description in which energy absorption occurs perturbatively. We implement this procedure using the van Vleck inverse-frequency expansion~\cite{PhysRevB.84.235108,PhysRevB.82.235114,PhysRevLett.110.026603}, which provides a systematic perturbative construction of a time-independent Floquet Hamiltonian to leading order in $1/\omega$. Within this framework, the effective Hamiltonian is\begin{align}\label{eq_4}
			\mathcal{H}^{\rm eff}_{\rm d}(\mathbf{k}) \approx {} &\mathcal{H}^{\rm F}_0(\mathbf{k})
			+ \frac{1}{\hbar \omega}
			\left[ \mathcal{H}^{\rm F}_{-1}(\mathbf{k}),\, \mathcal{H}^{\rm F}_{+1}(\mathbf{k}) \right] \notag \\ {} &\redd{+ \frac{1}{2 \hbar \omega}
			\left[ \mathcal{H}^{\rm F}_{-2}(\mathbf{k}),\, \mathcal{H}^{\rm F}_{+2}(\mathbf{k}) \right]},
		\end{align}where the Fourier components~\redd{($m = 0, \pm 1, \pm 2$)} of the driven Hamiltonian are defined as\begin{equation}
			\mathcal{H}^{\rm F}_m(\mathbf{k})
			= \frac{\omega}{2\pi}\int_0^{\frac{2\pi}{\omega}} dt\;
			H_{\rm d}\big(\mathbf{k}+e\mathbf{A}(t)/\hbar\big)\, e^{i m \omega t}.
		\end{equation}The resulting effective Hamiltonian governs the slow, stroboscopic dynamics of the driven system and faithfully captures its long-time behavior after averaging over rapid intra-period oscillations.
		
		Substituting $\mathbf{k} \to \mathbf{k} + e\mathbf{A}(t)/\hbar$ in $H_{\mathrm{d}}(\mathbf{k})$ via the minimal coupling scheme expands the Hamiltonian in powers of $\mathbf{A}$. \iffalse leading to} \begin{align}
			H_\text{d}(\mathbf{k}+\mathbf{A}(t)) = {} &H_\text{d}(\mathbf{k})  +\redd{\frac{e^2}{m_{\rm e}}}M_{S}^{\rm d}\sigma _{z}A_{x}(t)A_{y}(t) \nonumber \\
			{} &+\left[ \left(2 \redd{\frac{e}{\hbar}}\varepsilon\sigma _{0}+\redd{\frac{e \hbar}{m_{\rm e}}}M_{C}^{\rm d}\sigma _{z}\right) k_{x}\ +
			\redd{\frac{e}{\hbar}}\lambda _{\rm R}\sigma _{y} +\redd{\frac{e\hbar }{m_{\rm e}}}M_{S}^{\rm d}\sigma
			_{z}k_{y}\right] A_{x}(t) \nonumber \\
			{} &+\left[ \left(2 \redd{\frac{e}{\hbar}}\varepsilon\sigma _{0}-\redd{\frac{e \hbar}{m_{\rm e}}}M_{C}^{\rm d}\sigma _{z}\right) k_{y}
			-\redd{\frac{e}{\hbar}}\lambda _{\rm R}\sigma _{x} \ +\redd{\frac{e\hbar }{m_{\rm e}}}M_{S}^{\rm d}\sigma
			_{z}k_{x}\right] A_{y}(t)  \notag \\ {}&+\left( \redd{\frac{e^2}{\hbar^2}}\varepsilon\sigma _{0}+\redd{\frac{e^2}{m_{\rm e}}}\frac{M_{C}^{\rm d}}{2}\sigma _{z}\right)
			A_{x}^{2}(t)+\left( \redd{\frac{e^2}{\hbar^2}}\varepsilon\sigma _{0}-\redd{\frac{e^2}{m_{\rm e}}}\frac{M_{C}^{\rm d}}{2}\sigma _{z}\right)
			A_{y}^{2}(t)\,.
		\end{align}\fi For notational convenience, we introduce the coefficients \begin{align}
			M_{C}^{\rm d} \equiv M_{\rm d}\cos\left(2\theta_{M}^{\rm d}\right)\,,\qquad M_{S}^{\rm d} \equiv M_{\rm d}\sin\left(2\theta_{M}^{\rm d}\right)\,.
		\end{align}\redd{For the zeroth harmonic ($m = 0$), we obtain
	\begin{subequations}
				\begin{align}
					\frac{\omega}{2\pi}\int_0^{2\pi/\omega} dt\, A_x(t) A_y(t) 
					&= 0, \\
					\frac{\omega}{2\pi}\int_0^{2\pi/\omega} dt\, A_{x/y}(t) 
					&= 0, \\
					\frac{\omega}{2\pi}\int_0^{2\pi/\omega} dt\, A_{x/y}^2(t) 
					&= \frac{A_0^2}{2} \bigl(1 + S^2\bigr).
				\end{align}
		\end{subequations}For the first harmonics ($m = \pm 1$), we find\begin{subequations}
		\begin{align}
			\frac{\omega}{2\pi}\int_0^{2\pi/\omega} A_x(t) A_y(t) \, e^{\pm i \omega t}\, dt 
			&= \frac{\pm i A_0^2 \mathcal{S} (\eta_2 - \eta_1)}{4} e^{\mp i\phi}, \\
			\frac{\omega}{2\pi}\int_0^{2\pi/\omega} A_x(t) \, e^{\pm i \omega t}\, dt 
			&= \frac{A_0}{2}, \\
			\frac{\omega}{2\pi}\int_0^{2\pi/\omega} A_y(t) \, e^{\pm i \omega t}\, dt 
			&= \frac{\pm i A_0 \eta_1}{2}, \\
			\frac{\omega}{2\pi}\int_0^{2\pi/\omega} A_x^2(t) \, e^{\pm i \omega t}\, dt 
			&= \frac{A_0^2 \mathcal{S}}{2} e^{\mp i\phi}, \\
			\frac{\omega}{2\pi}\int_0^{2\pi/\omega} A_y^2(t) \, e^{\pm i \omega t}\, dt 
			&= \frac{A_0^2 \mathcal{S} \eta_1 \eta_2}{2} e^{\mp i\phi}.
		\end{align}
	\end{subequations}Finally, for the second harmonics ($m = \pm 2$), we obtain\begin{subequations}
	\begin{align}
		\frac{\omega}{2\pi}\int_0^{2\pi/\omega} A_x(t) A_y(t) \, e^{\pm 2 i \omega t}\, dt 
		&= \pm \frac{i A_0^2}{4}
		\eta_1 , \\
		\frac{\omega}{2\pi}\int_0^{2\pi/\omega} A_x(t) \, e^{\pm 2 i \omega t}\, dt 
		&= \frac{A_0 \mathcal{S}}{2} e^{\mp i\phi}, \\
		\frac{\omega}{2\pi}\int_0^{2\pi/\omega} A_y(t) \, e^{\pm 2 i \omega t}\, dt 
		&= \frac{\pm i A_0 \eta_2 \mathcal{S}}{2} e^{\mp i\phi}, \\
		\frac{\omega}{2\pi}\int_0^{2\pi/\omega} A_x^2(t) \, e^{\pm 2 i \omega t}\, dt 
		&= \frac{A_0^2}{4}, \\
		\frac{\omega}{2\pi}\int_0^{2\pi/\omega} A_y^2(t) \, e^{\pm 2 i \omega t}\, dt 
		&= -\frac{A_0^2}{4}.
	\end{align}
\end{subequations}}
		
		\redd{As a result, we obtain \begin{align}
				\mathcal{H}^{\mathrm{F}}_m(\mathbf{k}) 
				&= h_0^{(m)}(\mathbf{k})\,\sigma_0 
				+ h_x^{(m)}\,\sigma_x 
				+ h_y^{(m)}\,\sigma_y 
				+ h_z^{(m)}(\mathbf{k})\,\sigma_z\, ,
		\end{align}with\begin{subequations}\label{eq_11}
		\begin{align}
			h_0^{(0)}(\mathbf{k}) = {} &\frac{A_0^2 e^2 \varepsilon \left( 1 + S^2  \right)}{\hbar^2} ,\\
			h_x^{(0)}(\mathbf{k}) = {} &0\,,\\
			h_y^{(0)}(\mathbf{k}) = {} & 0\,,\\
			h_z^{(0)}(\mathbf{k}) = {} & 0\,,\\
			h_0^{(\pm 1)}(\mathbf{k}) = {} & \frac{A_0}{2 \hbar^2} \Big[2  e \varepsilon \hbar \left(k_x\pm  i \eta_1 k_y\right) \notag \\ &{}+ A_0 S e^2 \varepsilon e^{\mp i \phi} \left(  1 + \eta_1 \eta_2\right)\Big]\, ,\\
			h_x^{(\pm 1)}(\mathbf{k}) = {} & \mp\frac{ i A_0 e \eta_1 \lambda_{\rm R}}{2 \hbar}\, ,\\
			h_y^{(\pm 1)}(\mathbf{k}) = {} & \frac{A_0 e \lambda_{\rm R}}{2 \hbar}
			\, ,\\
			h_z^{(\pm 1)}(\mathbf{k}) = {} & \frac{A_0 e}{4 m_e} e^{\mp i\phi} \Biggl[
			2\hbar e^{\pm i\phi} (M_{\rm C}^{\rm d} k_x + M_{\rm S}^{\rm d} k_y) 
			\notag \\ &{}\mp  2i \eta_1 \hbar e^{\pm i\phi} (M_{\rm C}^{\rm d} k_y - M_{\rm S}^{\rm d} k_x) 
			\notag \\ {} &+ A_0 \mathcal{S} e \Bigl[ 
			M_{\rm C}^{\rm d} (1 - \eta_1\eta_2) 
			\pm i M_{\rm S}^{\rm d} (\eta_2 - \eta_1)
			\Bigr]
			\Biggr]\,,\\
			h_0^{(\pm 2)}(\mathbf{k}) = {} & \frac{A_0 S e \varepsilon \left(k_x\pm i  \eta_2 k_y\right) e^{\mp i \phi}}{\hbar}
			\, ,\\
			h_x^{(\pm 2)}(\mathbf{k}) = {} &\mp  \frac{i  A_0 S e \eta_2 \lambda_{\rm R}  \, e^{\mp i \phi}}{2 \hbar}
			 \, ,\\
			h_y^{(\pm 2)}(\mathbf{k}) = {} & \frac{e \lambda_{\rm R} A_0 \mathcal{S}}{2 \hbar}e^{\mp i\phi} \, ,\\
			h_z^{(\pm 2)}(\mathbf{k}) = {} & \frac{A_0 e}{4 m_{\rm e}} e^{\mp i \phi} \Bigg( A_0  e e^{\pm i \phi} \left(M_{\rm C}^{\rm d} \pm i M_{\rm S}^{\rm d} \eta_1 \right) \notag \\ &{}+ 2 \mathcal{S}\hbar \bigg(M_{\rm C}^{\rm d} k_x + M_{\rm S}^{\rm d} k_y \notag \\ &{}\pm  i \eta_2\big(M_{\rm S}^{\rm d}k_x - M_{\rm C}^{\rm d} k_y\big)\bigg)\Bigg)
			 \, .
		\end{align}
	\end{subequations}Next, we compute the commutators $\frac{1}{\hbar \omega} \left[ \mathcal{H}^{\rm F}_{-1}(\mathbf{k}), \, \mathcal{H}^{\rm F}_{+1}(\mathbf{k}) \right]$ and $\frac{1}{2\hbar \omega} \left[ \mathcal{H}^{\rm F}_{-2}(\mathbf{k}), \, \mathcal{H}^{\rm F}_{+2}(\mathbf{k}) \right]$. For a general \(\mathbf{h} \cdot \boldsymbol{\sigma}\), we know that $\left[ \mathbf{h}_1 \cdot \boldsymbol{\sigma}, \, \mathbf{h}_2 \cdot \boldsymbol{\sigma} \right] = 2 i \, (\mathbf{h}_1 \times \mathbf{h}_2) \cdot \boldsymbol{\sigma}$. Moreover, any \(\sigma_0\) term commutes with everything, so $\left[ h_0^{(-1)} \, \sigma_0, \, \cdots \right] = 0$ and $\left[ \cdots, \, h_0^{(+1)} \, \sigma_0 \right] = 0$. Thus, the commutator reduces to\begin{align}
		\frac{1}{\hbar \omega} \left[ \mathcal{H}^{\rm F}_{-m}(\mathbf{k}), \, \mathcal{H}^{\rm F}_{+m}(\mathbf{k}) \right] = \frac{2 i}{\hbar \omega} \left( \mathbf{h}^{(-m)} \times \mathbf{h}^{(+m)} \right) \cdot \boldsymbol{\sigma}\, ,
\end{align}with cross product formula\begin{align}
\mathbf{h}^{(-m)} \times \mathbf{h}^{(+m)} 
&=
\begin{pmatrix}
	h_y^{(-m)} h_z^{(+m)} - h_z^{(-m)} h_y^{(+m)} \\\\
	h_z^{(-m)} h_x^{(+m)} - h_x^{(-m)} h_z^{(+m)} \\\\
	h_x^{(-m)} h_y^{(+m)} - h_y^{(-m)} h_x^{(+m)}
\end{pmatrix}.
\end{align}Using the expressions in Eq.~\eqref{eq_11}, each component can be found. }Finally, we obtain\begin{align}\label{eq_8}
			H_{\mathrm{d}}^{\rm eff}(\textbf{k}) ={}&
			\Big[\varepsilon\left(k_x^2 + k_y^2\right) + \mu_\omega \Big]\sigma_0
			+ \Big[(\lambda_{\rm R} + \lambda_{{\rm R}\omega})k_x \notag \\ {} &+ \lambda_{{\rm D}\omega}k_y \Big]\sigma_y
			- \Big[(\lambda_{\rm R} - \lambda_{{\rm R}\omega})k_y + \lambda_{{\rm D}\omega}k_x \Big]\sigma_x	\notag \\ {} &+ M_\textbf{k}^{\rm d} \sigma_z
		+ \boldsymbol{\Omega}\cdot\boldsymbol{\sigma},
			\end{align}with\begin{subequations}
			\begin{align}
				\mu_\omega &= \redd{\frac{e^2}{\hbar^2}}\varepsilon A_0^2\left(1 + \mathcal{S}^2\right), \\[6pt]
				\lambda_{{\rm R}\omega} &=  \redd{\frac{e^2}{m_{\rm e}}}\lambda_{\rm R} M_{\rm C}^{\rm d}  A_0^2 \frac{\redd{2\eta_1}+ \eta_2 \mathcal{S}^2}{\redd{2\hbar\omega}}, \\[6pt]
				\lambda_{{\rm D}\omega} &= \redd{\frac{e^2}{m_{\rm e}}}\lambda_{\rm R} M_{\rm S}^{\rm d} A_0^2  \frac{\redd{2\eta_1}+ \eta_2 \mathcal{S}^2}{\redd{2\hbar\omega}},\\
				\Omega_x &=\redd{\frac{e^3}{m_{\rm e}}}
				\frac{\lambda_{\rm R}}{\redd{4\hbar^2\omega}} A_0^3 \mathcal{S}
				\Big[\redd{(1-2 \eta_1\eta_2)}M_{\rm C}^{\rm d} \sin\left(\phi\right) 
				\notag \\ {} &\hspace*{2.5cm}+\redd{(\eta_1- 2\eta_2)} M_{\rm S}^{\rm d}\cos\left(\phi\right)
				\Big] , \\[8pt]
				\Omega_y &=\redd{\frac{e^3}{m_{\rm e}}}
				\frac{\lambda_{\rm R}}{\redd{4\hbar^2\omega}} A_0^3 \mathcal{S}
				\Big[
				\redd{(\eta_1 \eta_2 - 2)}M_{\rm S}^{\rm d}\sin\left(\phi\right)\notag \\ {} &\hspace*{2.5cm}+\redd{(2\eta_1-\eta_2)} M_{\rm C}^{\rm d}\cos\left(\phi\right)
				\Big] , \\[8pt]
				\Omega_z &=\redd{-\frac{e^2}{\hbar^2}}
				\lambda_{\rm R}^2 A_0^2 \frac{\redd{2\eta_1}+ \eta_2 \mathcal{S}^2}{\redd{2\hbar\omega}} .
			\end{align}
			\end{subequations}The term proportional to $A_0^2(1+\mathcal{S}^2)\sigma_0$ represents a light-induced scalar (AC Stark) shift arising from second-order absorption--emission processes that rigidly shift the electronic spectrum. \redd{The term $\mathbf{h}^{(-m)} \times \mathbf{h}^{(+m)} $ captures virtual single- and second-photon processes at the fundamental frequency $\omega$ and $2\omega$}, renormalizing the RSOC and momentum-dependent altermagnetic exchange. In addition to the \(\mathcal{S}\)-dependent contributions in the AC Stark and Rashba terms, these latter commutators emerge specifically under the two-color driving protocol. The relative phase $\phi$ and amplitude ratio $\mathcal{S}$ control constructive or destructive interference between these pathways, enabling selective enhancement or suppression of spin-dependent Floquet band hybridization. 
			
			Upon rewriting the effective Hamiltonian in the form
			$H_{\mathrm{d}}^{\rm eff}(\mathbf{k}) = h_0(\mathbf{k})\,\sigma_0 + \boldsymbol{h}(\mathbf{k})\cdot\boldsymbol{\sigma}$,
			the eigenvalues are readily obtained, giving rise to the upper and lower energy bands\begin{align}\label{eq_9}
				E_\pm(\mathbf{k}) = h_0(\mathbf{k}) \pm |\boldsymbol{h}(\mathbf{k})| \, .
			\end{align}In the limit $\mathcal{S}=0$, the model reduces to monochromatic circularly polarized light~\cite{k3xb-8pts,tm58-lbdl,Fu2025Floquet,Fu2025Light}. The momentum-linear terms $\sigma_x$ and $\sigma_y$ encode Floquet-modified spin--momentum locking: Rashba- and Dresselhaus-type terms, $\lambda_{{\rm R}\omega}$ and $\lambda_{{\rm D}\omega}$, arise from virtual single-photon processes that anisotropically renormalize the Rashba interaction in $d_{x^2-y^2}$ and $d_{xy}$ altermagnets, respectively. Higher-order multiphoton processes produce an effective Zeeman field $\boldsymbol{\Omega}\cdot\boldsymbol{\sigma}$, breaking time-reversal symmetry and generating momentum-independent spin splitting. These Floquet-induced terms vanish when $M_{\rm C}^{\rm d}= 0 $ and $M_{\rm S}^{\rm d}=0$, highlighting that such light responses are unique to altermagnets. In contrast, monochromatic circular driving preserves space--time rotational symmetry, forbidding static in-plane Zeeman fields $\Omega_x\sigma_x$ and $\Omega_y\sigma_y$. Among the Floquet-induced terms, $\lambda_{{\rm R}\omega}$, $\lambda_{{\rm D}\omega}$, and $\Omega_{x,y}$ play a key role in reshaping the Fermi surface, suppressing static spin susceptibility, and enhancing switching efficiency by activating additional RKKY channels.\begin{table}[t]
			\centering
			\resizebox{1\linewidth}{!}{\begin{tabular}{l|c|c|c|c|c}
					\hline\hline
					Term & One-color & Two-color & W/O $M_{\rm C}^{\rm d}$ 
					& W/O $M_{\rm S}^{\rm d}$ 
					& W/O altermagnetism?\\
					\hline
					$\mu_\omega$ 
					& \cmark (strong field) 
					& \cmark (weak fields) 
					& \cmark & \cmark & \cmark 
					\\[4pt]
					
					$\lambda_{{\rm R}\omega}$ 
					& \cmark (strong field) 
					& \cmark (weak fields) 
					& \cmark & \xmark & \xmark 
					\\[4pt]
					
					$\lambda_{{\rm D}\omega}$ 
					& \cmark (strong field) 
					& \cmark (weak fields) 
					& \xmark & \cmark & \xmark 
					\\[4pt]
					
					$\Omega_x$ 
					& \xmark
					& \cmark 
					& \xmark & \xmark & \xmark 
					\\[4pt]
					
					$\Omega_y$ 
					& \xmark 
					& \cmark
					& \xmark & \xmark & \xmark 
					\\[4pt]
					$\Omega_z$ 
					& \cmark (strong field) 
					& \cmark (weak fields) 
					& \cmark & \cmark & \cmark 
					\\
					\hline\hline
			\end{tabular}}
			\caption{Comparison of effective Floquet-induced terms and their efficiency for tuning the band structure under one-color and two-color laser driving. The last three columns summarize the dependence of Floquet-induced terms on the altermagnetic order parameters
				$M_{\rm C}^{\rm d}$ and $M_{\rm S}^{\rm d}$ to see how Hamiltonian terms appear without (W/O) altermagnetism.}	\label{tab1}
				\end{table}
			
			Table~\ref{tab1} compares one- and two-color driving protocols. Monochromatic driving only provides rigid band shifts and magnetizations, which compete with Rashba and altermagnetic splittings and thus require strong fields to strongly reshape the Fermi surface for a Lifshitz-like transition. Bichromatic driving, in contrast, enables similar band-structure tuning at much lower peak intensities, making Floquet engineering both robust and experimentally practical. In particular, the decisive control channels $\Omega_{x,y}$ are absent in monochromatic driving and vanish without altermagnetic order. The last three columns show that while the scalar shift $\mu_\omega$ and out-of-plane Zeeman-like field $\Omega_z$ persist without altermagnetism, all in-plane spin--orbit renormalizations and Zeeman-like fields require altermagnetic order, i.e., finite $M_{\rm C}^{\rm d}$ and/or $M_{\rm S}^{\rm d}$.  
			
			Although in-plane Zeeman terms can arise from substrate-induced proximity effects~\cite{PhysRevLett.124.227001,PhysRevB.99.174511,PhysRevX.7.021032,HUANG2022488}, such approaches rely on static, material-specific interfaces that are fixed once fabricated. By contrast, two-color laser driving generates effective in-plane Zeeman-like fields intrinsically within the 2D Rashba altermagnet, allowing continuous, reversible, and symmetry-selective control via intensity, polarization, and relative phase. \red{This optical route also avoids} interface-induced disorder, uncontrolled hybridization, and additional symmetry breaking, providing a clean and versatile platform for engineering spin-dependent band structure effects beyond static heterostructures.

            We have shown that the host electrons in Rashba altermagnets exhibit novel spin-dependent behavior under a two-color laser driving scheme. To understand how these emergent features manifest in realistic experimental conditions—where materials inevitably contain dilute impurities (both charged and magnetic)—we focus here on magnetic impurities, as the driven host system develops intrinsically spin-dependent electronic responses. Magnetic impurities act as localized spin probes that couple to the itinerant electrons via a local exchange interaction. In contrast to extended magnetic order, a dilute impurity carries a well-defined localized moment, enabling a controlled analysis of how the host electronic system mediates spin-dependent interactions. Consequently, magnetic impurities serve as tunable test spins whose effective coupling directly encodes the spin susceptibility and Fermi-surface properties of the underlying material. Studying their interaction therefore provides a minimal and direct framework for characterizing carrier-mediated magnetism, spin anisotropy, and the impact of spin–orbit coupling and external driving on the material structure. Moreover, a single impurity only induces a local spin polarization of the conduction electrons, whereas the effective magnetic coupling emerges when the spin density generated by one impurity interacts with a second spatially separated impurity. The two-impurity configuration, therefore, represents the minimal and most transparent setup to extract the carrier-mediated exchange interaction and to analyze how its range, oscillation period, and anisotropy are modified by the Floquet-renormalized Rashba–alternagnetic band structure.
			
			\section{Magnetic impurities and the RKKY interaction}\label{s3}
			We consider two localized magnetic moments, $\mathbf{S}_1$ and $\mathbf{S}_2$, as shown in Fig.~\ref{f1}. They are separated by the vector $\mathbf{R}$ and the relative angle $\varphi_R$. The impurity spins are coupled to the itinerant electrons via a local exchange interaction\begin{align}
				H_{\mathrm{sd}}
				=
				J_{\mathrm{sd}}
				\sum_{i=1,2}
				\mathbf{S}_i \cdot \mathbf{s}(\mathbf{R}_i),
			\end{align}where $\mathbf{s}(\mathbf{r})
			=
			\frac{1}{2}
			c^{\dagger}(\mathbf{r})
			\boldsymbol{\sigma}
			c(\mathbf{r})$ is the electron spin density and $J_{\mathrm{sd}}$ denotes the exchange coupling strength, typically lying in the range $10~\text{meV}$ to $1~\text{eV}$ in metallic and
			semiconducting hosts~\cite{PhysRev.185.847,PhysRevB.69.121303,PhysRevB.87.045422}. In the weak-coupling regime, the interaction between impurities arises at second order in $J_{\mathrm{sd}}$ upon integrating out the itinerant electrons, yielding the RKKY interaction. Since the electronic system is periodically driven, the exchange interaction is evaluated using the static component of the Rashba-Floquet Green’s function derived from the effective Hamiltonian. The effective RKKY Hamiltonian takes the most general bilinear form\begin{subequations}
			\begin{align}
				H_{\mathrm{RKKY}}
				={} &
				\sum_{\alpha,\beta=x,y,z}
				J_{\alpha\beta}(\mathbf{R})
				S_1^{\alpha} S_2^{\beta}\,,\\
				 J_{\alpha\beta}(\mathbf{R})
				={} &
				- J_{\mathrm{sd}}^2
				\chi_{\alpha\beta}(\mathbf{R}).
			\end{align}\end{subequations}The static spin susceptibility of the host material is expressed as{\small\begin{equation}
					\chi_{\alpha\beta}(\mathbf{R})
					=
					-\frac{1}{\pi}
					\int_{-\infty}^\mu
					dE\,
					\mathrm{Im}
					\left[
					\mathrm{Tr}
					\left(
					\sigma_\alpha
					G(\mathbf{R},E)
					\sigma_\beta
					G(-\mathbf{R},E)
					\right)
					\right],
			\end{equation}}where\rd{\begin{align}\label{eq_21}
			G(\mathbf{R},E)
			= {} &
			\int
			\frac{d^2\mathbf{k}}{(2\pi)^2}
			\frac{e^{i\mathbf{k}\cdot\mathbf{R}}}
			{E - H_{\mathrm{d}}^{\rm eff}(\textbf{k}) + i\Gamma}
			\end{align}}is the retarded Floquet Green’s function of the Rashba altermagnet in real space, and $\mu$ denotes the Fermi energy, which coincides with the chemical potential in the zero temperature limit assumed throughout this work. \rd{Here, $\Gamma$ is a phenomenological broadening term representing finite quasiparticle lifetimes. In our periodically driven system, we assume the 2D Rashba altermagnet is coupled to a macroscopic fermionic reservoir (e.g., a substrate) that rapidly relaxes the electrons, thereby fixing the chemical potential $\mu$ and maintaining a steady-state electronic distribution that closely approximates the zero-temperature Fermi-Dirac distribution. This hierarchy of timescales—where the driving period is much shorter than the reservoir-induced relaxation time—ensures that the Floquet bands dictate the susceptibility while the bath dictates the occupations.} The $H_{\mathrm{d}}^{\rm eff}(\textbf{k})$ term is the Rashba-Floquet altermagnet Hamiltonian in Eq.~\eqref{eq_8}. The long-range behavior of the exchange interactions is governed by the poles and branch cuts of the Green’s function near the Fermi surface. Because the Floquet-renormalized quasiparticle spectrum determines these singularities, the resulting exchange coupling directly reflects the spin-split and anisotropic structure of the Rashba–alternating magnetic Fermi surface.
			
			In Rashba altermagnets, the matrix structure of Green’s function in spin space is due to the combined presence of RSOC and altermagnetic order, which break inversion and spin-rotation symmetries. As a result, all components of the exchange tensor $J_{\alpha\beta}(\mathbf{R})$ are, in general, finite. Writing the interaction explicitly, the RKKY Hamiltonian reads\begin{align}
				H_{\rm{RKKY}}
				= {} &
				J_{xx} S_1^{x}  S_2^{x}
				+
				J_{yy} S_1^{y}  S_2^{y}
				+
				J_{zz} S_1^{z}  S_2^{z}
				\notag \\ {} &+
				J_{xy}
				\left(
				S_1^{x} S_2^{y}
				+
				S_1^{y} S_2^{x}
				\right)
				+
				J_{xz}
				\left(
				S_1^{x} S_2^{z}
				+
				S_1^{z} S_2^{x}
				\right)
			\notag\\
			{} &	+
				J_{yz}
				\left(
				S_1^{y} S_2^{z}
				+
				S_1^{z} S_2^{y}
				\right)
				+
				\mathbf{D}(\mathbf{R}) \cdot
				\left(
				\mathbf{S}_1 \times \mathbf{S}_2
				\right),
			\end{align}where the Dzyaloshinskii--Moriya~(DM) vector is given by $\textbf{D} = (J_{yz} - J_{zy}, J_{zx} - J_{xz}, J_{xy} - J_{yx})$. The diagonal terms $J_{xx}$, $J_{yy}$, and $J_{zz}$ describe anisotropic Heisenberg exchange interactions. Their anisotropy reflects the momentum-dependent spin polarization of the Rashba-split altermagnetic bands and leads to direction-dependent oscillation periods and decay lengths of the RKKY interaction. Unlike conventional metals, where rotational symmetry enforces isotropic exchange, Rashba altermagnets generically exhibit $J_{xx} \neq J_{yy} \neq J_{zz}.$ The symmetric off-diagonal terms $J_{xy}$, $J_{xz}$, and $J_{yz}$ arise from spin--orbit--induced spin mixing in the electronic Green’s functions. These terms vanish in centrosymmetric or spin-rotation-invariant systems but are intrinsic to Rashba altermagnets. They encode couplings between orthogonal spin components of the impurities and favor noncollinear magnetic configurations. The antisymmetric part of the exchange tensor generates the DM interaction, which is permitted by the absence of inversion symmetry. Using $G(\textbf{R},E) = G_0(\textbf{R},E) \sigma_0 + \textbf{G}(\textbf{R},E) \cdot\boldsymbol{\sigma},$ the exchange interactions are given by
			
			{\small\begin{subequations}\label{eq_13}
					\begin{align}
						&\frac{J_{\alpha\alpha}(\mathbf{R})}{J_{\mathrm{sd}}^2}
						= {} -\frac{2}{\pi}\,
						\mathrm{Im}\!\int_{-\infty}^\mu \! dE \,
						\Big(
						G_0(\mathbf{R},E)G_0(-\mathbf{R},E)
						\notag \\ {} &+ G_\alpha(\mathbf{R},E)G_\alpha(-\mathbf{R},E)
						- \!\!\sum_{\beta\neq \alpha}
						G_\beta(\mathbf{R},E)G_\beta(-\mathbf{R},E)
						\Big)
						\, ,\\
						&\frac{J_{\alpha\beta}(\mathbf{R})}{J_{\mathrm{sd}}^2} = {} -\frac{ 2}{\pi} {\rm Im} \int^\mu_{-\infty} d \,E \,\Big(G_\alpha(\textbf{R},E)G_\beta(-\textbf{R},E)	\notag \\ {} & + G_\beta(\textbf{R},E)G_\alpha(-\textbf{R},E) + i \sum_\xi \varepsilon_{\alpha \beta \xi} [G_0(\textbf{R},E)G_\xi(-\textbf{R},E)\notag \\ {} & -  G_\xi(\textbf{R},E)G_0(-\textbf{R},E)]\Big)
						\,,\\
						&\frac{D_\alpha(\textbf{R})}{J_{\mathrm{sd}}^2} ={}  -\frac{4}{\pi}  {\rm Im} \int^\mu_{-\infty} d \,E \,i\Big(G_0(\textbf{R},E)G_\alpha(-\textbf{R},E)	\notag \\ {} &-G_\alpha(\textbf{R},E)G_0(-\textbf{R},E)\Big)\,.
					\end{align}
			\end{subequations}}
			
			Although Eqs.~(\ref{eq_13}) formally define the full tensorial RKKY interaction, closed analytical expressions are generally unattainable for the effective
			Hamiltonian in Eq.~(\ref{eq_8}). The Floquet-driven Rashba altermagnet hosts
			several competing energy scales, which together produce an anisotropic
			band structure lacking continuous rotational symmetry. Thus, the
			real-space Green’s functions involve 2D momentum integrals over
			nonseparable dispersions, rendering standard analytical treatments impractical
			beyond limiting cases. We therefore compute the Green’s functions numerically by
			discretizing the Brillouin zone on a dense $1000\times1000$ $\textbf{k}$-grid and
			performing the momentum and energy integrations up to the Fermi level at zero temperature. \red{Because the relevant physics is governed by low-energy states near the Fermi level and the drive frequency exceeds the electronic bandwidth, any non-periodicity at the Brillouin-zone boundaries does not affect the results, implying that Brillouin-zone edge states lie far from the low-energy window.}

            The diagonal exchange couplings are decomposed into an isotropic Heisenberg component, denoted as $J_{\rm H}=(J_{xx}+J_{yy}+J_{zz})/3$ and a uniaxial Ising anisotropy $J_{\rm I}=J_{zz}-(J_{xx}+J_{yy})/2$. This parametrization simplifies the presentation and interpretation of the RKKY results below. The antisymmetric exchange is captured by the DM vector $\mathbf{D}(\mathbf{R})$, whose magnitude sets the strength of the chiral coupling and whose orientation is dictated by RSOC and Floquet-induced symmetry breaking. The sign of each exchange term determines the preferred magnetic alignment of the impurities, while the magnitude controls the interaction strength. A positive (negative) Heisenberg coupling favors antiferromagnetic (AFM) or ferromagnetic (FM) alignment. The Ising term enforces collinear order along a selected axis, with its sign determining easy-axis behavior here. The antisymmetric DM interaction sets the chirality of noncollinear textures, selecting clockwise (CW) or counterclockwise (CCW) canting. In a nutshell, magnitudes determine interaction strength, while signs fix the magnetic ground state.

			\section{Results and discussion}\label{s4}
			Although the Fermi surface in the present model is anisotropic and band dependent, the electronic states relevant for the RKKY interaction are characterized by a typical momentum scale set by the Fermi momentum $k_{\rm F}$. We therefore measure all momenta in units of $k_{\rm F}=1$, without loss of generality. The dimensionless parameter $\mathcal{A}=eA_0/\hbar k_{\rm F}$ then directly quantifies the light-induced momentum shift relative to the Fermi surface due to the first beam, while $\mathcal{S}\mathcal{A}$ denotes the corresponding shift induced by the second beam. In dilute or low-carrier-density regimes, $k_{\rm F}$ is substantially reduced compared to conventional metals; for typical 2D densities $10^{10}$--$10^{12}\,\mathrm{cm}^{-2}$ one finds $k_{\rm F}\sim10^{7}$--$10^{8}\,\mathrm{m}^{-1}$. Requiring $eA_0/\hbar\sim k_{\rm F}$ then corresponds to electric-field strengths $E_0=\omega A_0\sim10^{7}$--$10^{8}\,\mathrm{V/m}$ for our representative driving frequency $\hbar\omega \geq 1\,\mathrm{eV}$, which aligns perfectly with visible-light fields, making these nonequilibrium phenomena accessible with standard optical setups. These values are well within the range of standard ultrafast optical experiments~\cite{Higuchi2017,doi:10.1126/science.1239834,McIver2020,PhysRevResearch.2.043408,liu2024evidencefloquetelectronicsteady} and lie far below strong-field or breakdown thresholds. Unless stated otherwise, all results are shown for scattering rate $\Gamma = 1$ meV.
			
			We also note that, because the continuum quadratic dispersion lacks particle--hole symmetry and the Floquet drive induces a scalar Stark shift, the chemical potential $\mu = 0$ considered throughout this manuscript (unless stated otherwise) does not correspond to half filling. Instead, the carrier density is controlled by the Fermi energy measured relative to the light-renormalized band minimum.
			
			While we present the RKKY results primarily for the $d_{x^2-y^2}$-wave altermagnet, the efficient optical control discussed below is not symmetry-specific. As demonstrated explicitly in the following band-structure analysis, the same mechanism operates in the $d_{xy}$-wave altermagnet, indicating that the results are generic across $d$-wave altermagnetic symmetries.\begin{figure}[t]
				\centering
				\includegraphics[width=0.95\linewidth]{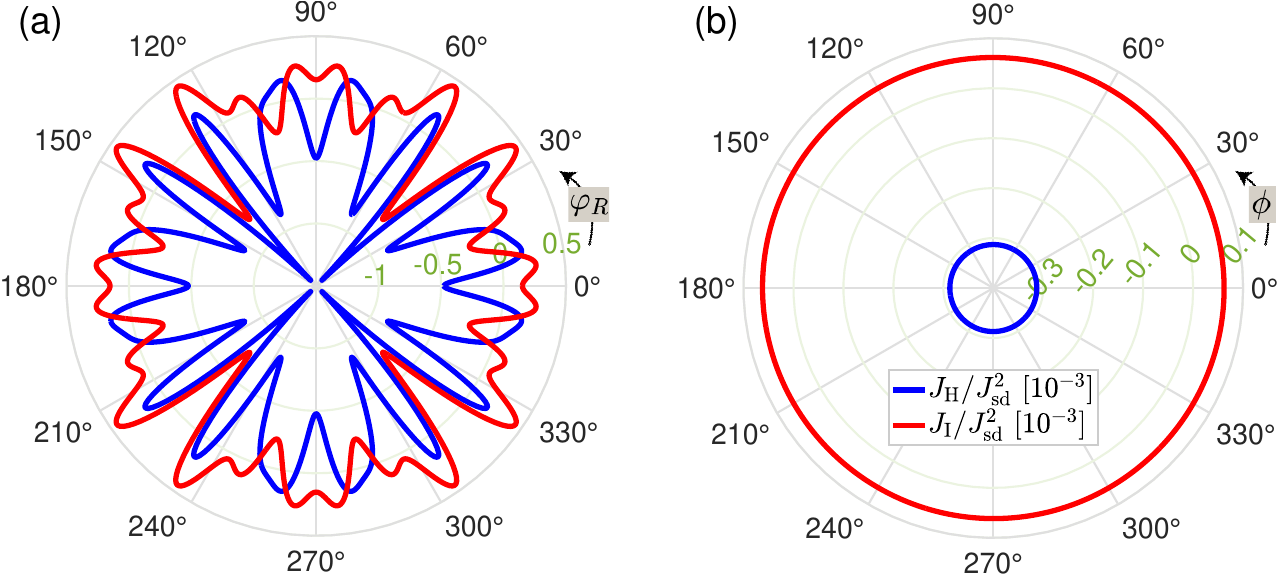}
				\caption{Angular dependence of the RKKY interaction in a 2D Rashba altermagnet for fixed impurity separation $R=35$~\AA, Rashba coupling $\lambda_{\rm R}=0.3~\mathrm{eV\cdot}$\AA, altermagnetic strength $M_{\rm d}=0.7$, and orientation $\theta_M^{\rm d}=0$. 
					(a) representative pristine Heisenberg ($J_{\rm H}$) and Ising ($J_{\rm I}$) components as functions of the relative impurity angle $\varphi_R$ in the absence of driving, illustrating the $2\pi$ periodicity and the intrinsic anisotropy of the altermagnetic band structure. 
					(b) Dependence of the same components on the relative angle $\phi$ between the two circularly polarized driving fields, showing an essentially isotropic response.}
				\label{f2n}
			\end{figure}
			
			The RKKY interaction is periodic in the relative impurity orientation,
			\begin{equation}
				H_{\rm RKKY}(\varphi_R+2\pi)=H_{\rm RKKY}(\varphi_R),
			\end{equation}as illustrated in Fig.~\ref{f2n}(a) for representative Heisenberg ($J_{\rm H}$) and Ising ($J_{\rm I}$) couplings in a pristine 2D Rashba altermagnet in the absence of driving. Since the driving fields do not couple directly to $\varphi_R$, varying this angle does not introduce qualitatively new Floquet-induced exchange mechanisms, but merely rescales the relative weights of existing components. We therefore fix $\varphi_R=\pi/3$ in the following.
			
			Furthermore, although the effective Rashba-Floquet Hamiltonian depends explicitly on the relative phase $\phi$ of the two laser fields through the in-plane Zeeman-like terms $\Omega_{x,y}$, the resulting static RKKY interaction is insensitive to $\phi$, as shown in Fig.~\ref{f2n}(b). Thus, varying the relative light angle does not introduce additional anisotropy, but instead leaves the symmetry of the Floquet-engineered exchange interactions intact. This follows from the inversion relation $G_\alpha(\mathbf{R},E)=G_\alpha^\ast(-\mathbf{R},E)$ [Eq.~\eqref{eq_13}]. Phase-dependent contributions enter with opposite signs at $\mathbf{R}$ and $-\mathbf{R}$, reflecting the inversion relation $G_\alpha(\mathbf{R},E)=G_\alpha^\ast(-\mathbf{R},E)$. As a result, all $\phi$-dependent terms appear in antisymmetric combinations that cancel when taking the imaginary part of the exchange integrals, which involve traces over products of Green's functions. We therefore fix $\phi=\pi/3$ throughout. Phase sensitivity may re-emerge for noncollinear impurities or in the presence of dynamical exchange, where the $\mathbf{R}\leftrightarrow-\mathbf{R}$ symmetry and the advanced–retarded cancellation no longer hold; these effects are beyond the scope of the present work. The difference in the radii of two RKKY components reflects distinct Floquet renormalizations of the corresponding exchange channels.\begin{figure}[t]
				\centering
				\includegraphics[width=0.95\linewidth]{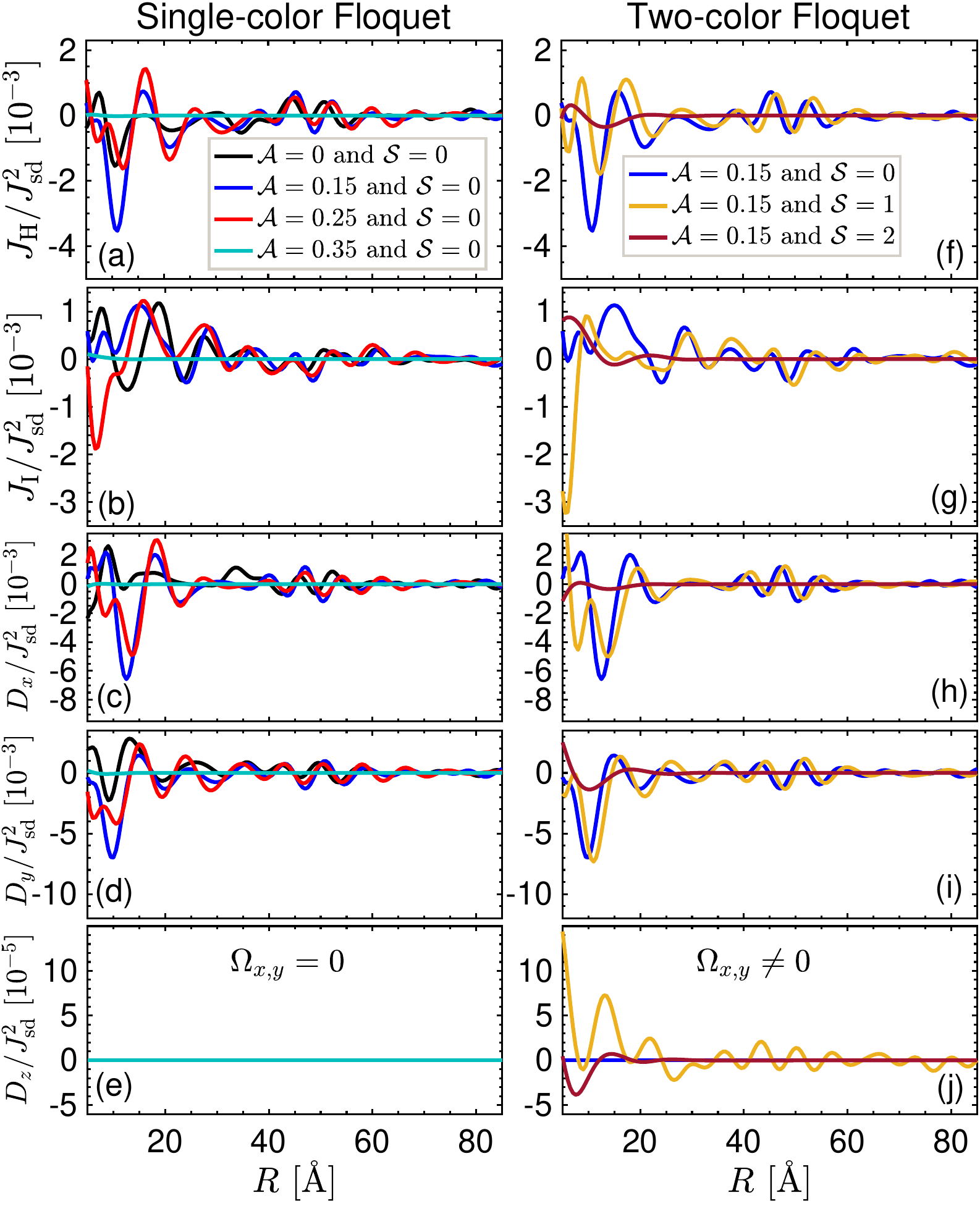}
				\caption{Distance dependence of the Heisenberg ($J_{\rm H}$), Ising ($J_{\rm I}$), and DM ($D_\alpha$) components of the RKKY interaction under single- and two-color driving for fixed Rashba coupling $\lambda_{\rm R}=0.3~\mathrm{eV\cdot}$\AA, altermagnetic strength $M_{\rm d}=0.7$, and orientation $\theta_M^{\rm d}=0$. While single-color driving [(a)--(e)] suppresses all exchange channels only at large amplitudes, two-color driving achieves comparable suppression at substantially reduced peak fields [(f)--(j)] with a 54\% reduction in the primary beam amplitude and the addition of a secondary beam whose amplitude is also reduced by 10\%, demonstrating a markedly more efficient control protocol.}
				\label{f2}
			\end{figure}
			
			Figure~\ref{f2} directly compares monochromatic and bichromatic Floquet control of the RKKY interaction in the \(d_{x^2-y^2}\)-wave Rashba altermagnet. Under single-color driving [Figs.~\ref{f2}(a)–(e)], all exchange channels are suppressed only once the drive amplitude exceeds a critical value $\mathcal{A}_{\rm c}\simeq0.33$ (see the videos in the Supplemental Material (SM)~\cite{SM}). In contrast, bichromatic driving [Figs.~\ref{f2}(f)–(j)] achieves an equivalent suppression using two weaker beams, with the primary amplitude fixed at $\mathcal{A}=0.15$, corresponding to a \(54\%\) reduction to \(\mathcal{A}_{\rm c}\), and the second beam providing $\mathcal{S}\mathcal{A} = 0.3$, corresponding to  \(\simeq 10\%\) reduction to \(\mathcal{A}_{\rm c}\). \rd{To properly compare the driving protocols, we evaluate the total optical intensity. Under single-color driving, switching occurs at a critical amplitude $\mathcal{A}_c \simeq 0.33$. Because the electric-field amplitudes of the fundamental and second-harmonic beams scale as $E_1 = \omega A_0$ and $E_2 = 2\omega \mathcal{S} A_0$, the total optical intensity scales approximately as $I_{\rm tot} \propto \mathcal{A}^2(1 + 4\mathcal{S}^2)$. Consequently, the monochromatic switching threshold requires an intensity proportional to $\sim 0.1089$. In the bichromatic scheme (e.g., $\mathcal{A} = 0.15$ and $\mathcal{S} = 1$), the required total intensity is $I_{\rm tot} \propto 0.1125$, and for higher $\mathcal{S}$, the intensity is indeed higher, indicating that the total injected optical power is comparable or higher. However, the critical experimental advantage of bichromatic driving lies in distributing the necessary field strength across two distinct frequency channels. By utilizing interference between the $\omega$ and $2\omega$ fields, the peak amplitude of the primary beam is reduced by roughly $54\%$ compared to $\mathcal{A}_c$. This redistribution allows for robust Floquet control and the initiation of the Lifshitz-like transition without subjecting the material to a single, highly intense fundamental beam, thereby minimizing the risk of optical breakdown and higher-order nonlinear damage at a single wavelength.}
			
			Under both driving protocols, the magnetic impurities exhibit separation-dependent FM/ or AFM and CW or CCW alignments, reflecting sign changes that evolve with the light intensity. While single-color driving suppresses the RKKY
			interaction in a largely monotonic fashion, at short distances, bichromatic driving induces a pronounced reshaping of the oscillatory structure of the exchange interaction. This behavior reflects interference between photon sectors, which modifies the real-space Green’s functions and selectively suppresses Fermi-surface contributions while leaving short-distance, high-momentum processes partially intact. This effect arises from Floquet-induced in-plane Zeeman-like fields $\Omega_{x,y}$ generated by two-color driving, as noted in Tab.~\ref{tab1}, which preferentially modify high-momentum states that dominate the short-range exchange.\begin{figure}[t]
				\centering
				\includegraphics[width=0.95\linewidth]{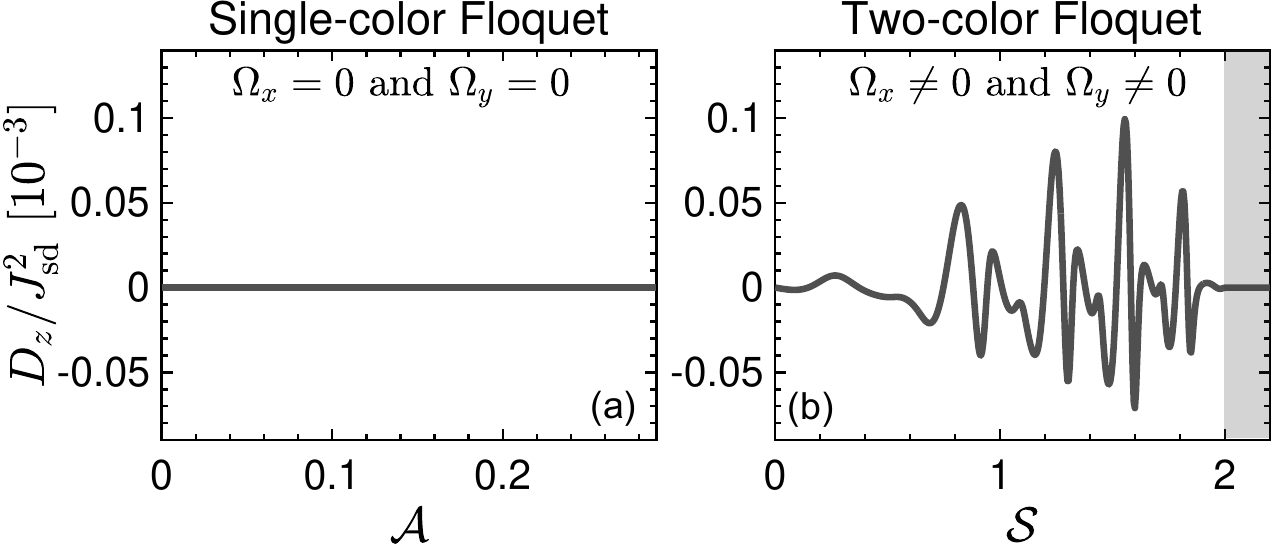}
				\caption{Out-of-plane DM interaction $D_z$ as a function of the drive parameter (a) $\mathcal{A}$ and (b) $\mathcal{S}$ under single- and two-color driving, respectively, at fixed impurity separation $R = 35$ \AA, Rashba coupling $\lambda_{\rm R}=0.3~\mathrm{eV\cdot}$\AA, altermagnetic strength $M_{\rm d}=0.7$, and altermagnetic orientation $\theta_M^{\rm d}=0$. While $D_z$ is strictly forbidden under monochromatic circular driving, bichromatic excitation induces a sizable and oscillatory $D_z$ due to Floquet-engineered in-plane Zeeman-like fields $\Omega_x$ and $\Omega_y$ in Eq.~\eqref{eq_8}.}
				\label{f3}
			\end{figure}
			
			A key qualitative distinction between the two driving protocols, Figs.~\ref{f2}(e) and \ref{f2}(f), is the emergence of a finite out-of-plane DM interaction $D_z$ under bichromatic excitation, as also shown in Fig.~\ref{f3}. This component is symmetry-forbidden in equilibrium and under single-color driving, and cannot be activated by simply increasing the drive amplitude $\mathcal{A}$ [Fig.~\ref{f3}(a)]. Microscopically, the in-plane Zeeman-like terms $\Omega_{x,y}$ generated by two-color interference enhance antisymmetric spin mixing in the real-space Green’s functions, directly producing a finite $D_z$ as a genuine dynamical Floquet effect. Moreover, $D_z$ vanishes permanently for $\mathcal{S} \gtrsim \mathcal{S}_{\rm c} \simeq 2$ (corresponding to $\mathcal{S}_{\rm c}\mathcal{A} \approx 0.3$) [Fig.~\ref{f3}(b)], as discussed below.\begin{figure*}[t]
				\centering
				\includegraphics[width=0.95\linewidth]{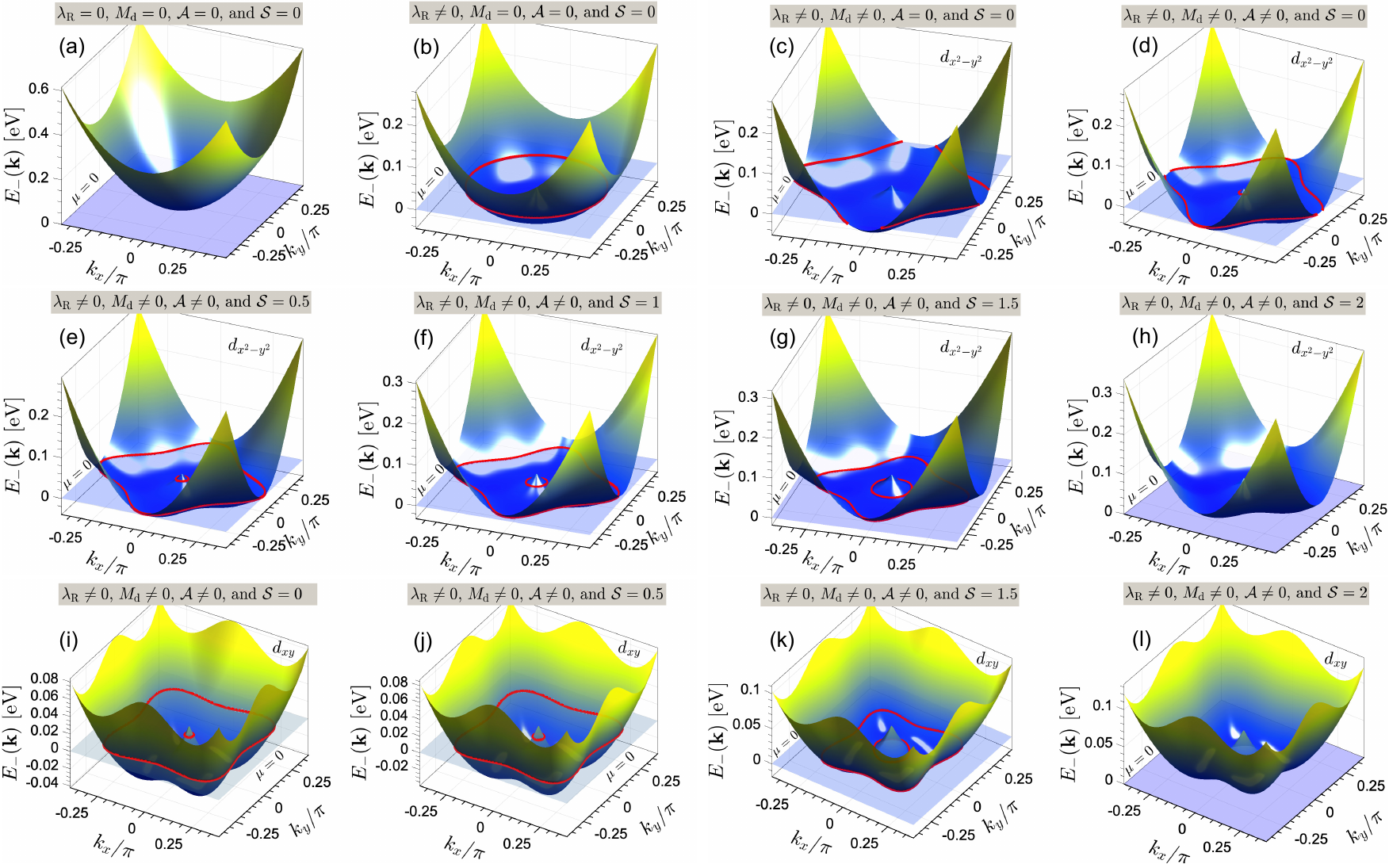}
				\caption{Evolution of the lower Floquet quasienergy band $E_-(\mathbf{k})$ in a 2D Rashba altermagnet under monochromatic and bichromatic driving for $d_{x^2-y^2}$- and $d_{xy}$-wave altermagnets. Starting from a (a) parabolic dispersion above the Fermi level ($\mu=0$) in the absence of RSOC, altermagnetism, and driving, successive tuning of (b) $\lambda_{\rm R}$, (c) $M_{\rm d}$, and (d)--(l) the laser parameters reshapes the low-energy electronic structure. Red lines on the $\mu=0$ plane indicate the corresponding Fermi surfaces. (d) Single-color driving ($\mathcal{A}=0.15$, $\mathcal{S}=0$) in the $d_{x^2-y^2}$-wave altermagnet modifies the band extrema via virtual photon processes, partially tuning states at the Fermi level. In contrast, (e)--(h) bichromatic driving systematically shifts the global minimum of $E_-(\mathbf{k})$ upward with increasing $\mathcal{S}$, inducing a controlled Lifshitz-like disappearance of the Fermi surface. (i)--(l) The same behavior is observed for the $d_{xy}$-wave altermagnet, demonstrating that the Floquet-engineered Lifshitz transition is robust across different altermagnetic symmetries. Parameters are $\lambda_{\rm R}=0.3~\mathrm{eV\cdot}$\AA, $M_{\rm d}=0.7$, $\eta_{1,2}=+1$, and $\theta_M^{\rm d}=0~(\pi/4)$ for the $d_{x^2-y^2}$- ($d_{xy}$-) wave case.}
				\label{f4}
			\end{figure*} 
			
			Practically, a finite out-of-plane component of the DM interaction, \(D_z\), favors in-plane cycloidal or helical spin spirals and is essential for stabilizing in-plane chiral textures such as N\'eel-type domain walls, planar spin spirals, and chiral edge states in 2D magnets~\cite{Niu2024,CAMLEY2023100605,Hoffmann2017,PhysRevLett.122.257202,PhysRevLett.128.167202,PhysRevB.102.024417,PhysRevB.110.014402}. In equilibrium Rashba altermagnets, as well as under single-color driving, crystalline symmetries typically forbid a finite \(D_z\), rendering these textures unstable or symmetry-forbidden. By contrast, within the Floquet framework, two-color laser driving dynamically breaks the relevant symmetries and induces a finite \(D_z\), thereby enlarging the magnetic phase space and enabling in-plane chiral states inaccessible under equilibrium or single-color conditions. Moreover, \(D_z\) enables robust information encoding and low-energy manipulation in racetrack and spintronic devices.
			
			\rd{To rigorously ensure the robustness of the proposed on-off switching mechanism against dissipation and nonthermal occupations, we have further evaluated the RKKY interaction in the presence of a phenomenological scattering rate $\Gamma$. As detailed in Appendix \ref{ap1}, while increased dissipation exponentially dampens the absolute magnitude of the exchange coupling and smooths high-frequency quantum interference oscillations, the critical on-off switching threshold ($\mathcal{S}_c \simeq 2$) remains robustly well-defined. This confirms that the macroscopic Floquet-induced band shift driving the Lifshitz-like transition survives realistic thermal and phenomenological smearing.}
			
			\rd{Because our proposed switching mechanism requires a relatively strong second-harmonic component ($\mathcal{S}_c \simeq 2$), it is crucial to ensure that higher-order multiphoton processes do not invalidate the truncated effective Hamiltonian. However, by operating in the high-frequency regime ($\hbar\omega \ge 1$ eV), the drive frequency remains much larger than the relevant low-energy electronic scales (Rashba coupling, altermagnetic exchange, and Fermi-level bandwidth). To definitively verify this, we benchmarked our analytical $1/\omega$ van Vleck expansion against an exact numerical Floquet-Sambe diagonalization~\cite{keliri2026sambeapproachfloquetlindbladopen} near the critical switching threshold. As detailed in Appendix \ref{ap2}, the analytical framework shows excellent quantitative agreement with the full numerical evaluation, confirming that the band-edge shift and the robust emergence of all relevant Fourier components are accurately captured.}
			
			The suppression of the RKKY interaction, beyond the critical values \(\mathcal{A}_{\rm c} \simeq 0.33\) and \(\mathcal{S}_{\rm c} \simeq 2\) for both the single- and two-color schemes, is directly linked to the Floquet-modified band structure.  The strength, sign, and anisotropy of the RKKY coupling are determined by the dispersion, spin texture, and density of states near the Fermi level. Accordingly, in Fig.~\ref{f4}, we focus on the lower Floquet band dispersion of the effective Hamiltonian in Eq.~\eqref{eq_8} for different parameters. Red lines in Fig.~\ref{f4} indicate the states available at the Fermi surface. Starting from a conventional 2D electron gas [Fig.~\ref{f4}(a)] with $M_{\rm d}=\lambda_{\rm R}=\mathcal{A}=\mathcal{S}=0$, the dispersion is purely parabolic and touches the chemical potential at $\mathbf{k}=0$, yielding a negligible density of states and RKKY interactions. Introducing RSOC [Fig.~\ref{f4}(b)] lowers the band minimum and generates finite Fermi-level states, while a strong altermagnetic order [Fig.~\ref{f4}(c)] renders the dispersion anisotropic with momentum-dependent spin splitting, thereby shaping the symmetry and anisotropy of the RKKY interaction.
			
\begin{figure}[t]
				\centering
				\includegraphics[width=0.95\linewidth]{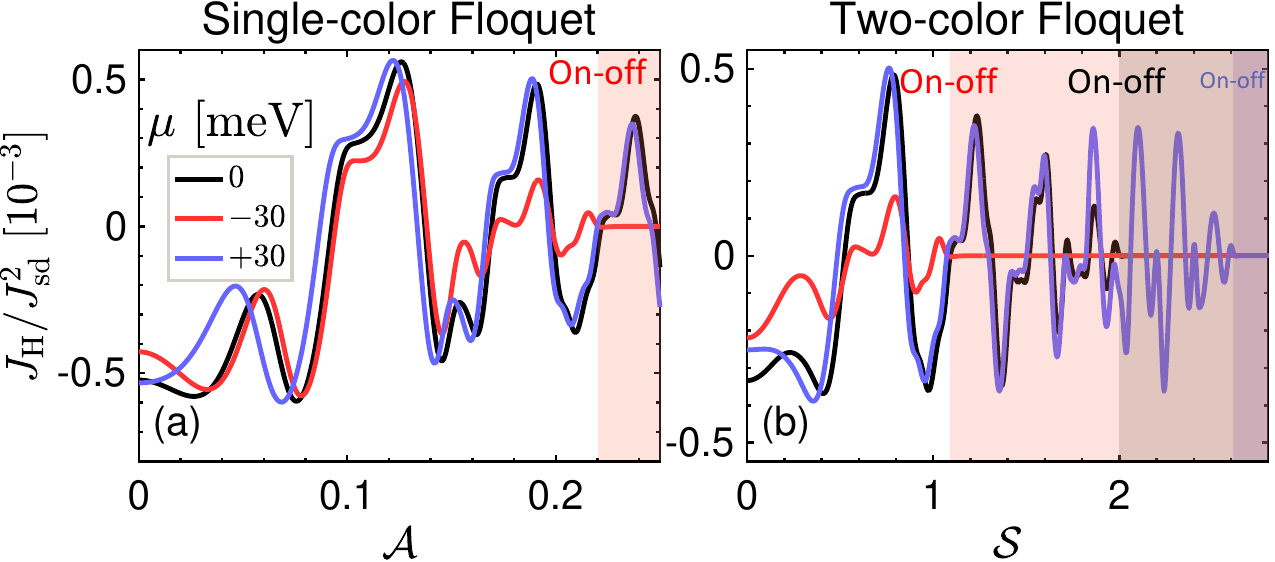}
				\caption{Dependence of the representative Heisenberg RKKY interaction $J_{\rm H}$ on mono- and bi-chromatic Floquet driving for different chemical potentials $\mu$: hole doping ($\mu<0$), charge neutrality ($\mu=0$), and electron doping ($\mu>0$). 
					(a) Single-color driving: $J_{\rm H}$ versus drive amplitude $\mathcal A$ at $\mathcal S=0$. 
					(b) Two-color driving: $J_{\rm H}$ versus relative strength $\mathcal S$ at $\mathcal A=0.15$. 
					Both protocols allow optical control and eventual suppression of the exchange, but the onset, robustness, and switching threshold are asymmetric in $\mu$: hole doping accelerates suppression, while electron doping enhances $J_{\rm H}$ and delays switching. Parameters: $\lambda_{\rm R} = 0.3~\mathrm{eV\cdot}$\AA, $M_{\rm d} = 0.7$, and $R = 35$~\AA.
				}
				\label{f7}
			\end{figure}

			Under single-color circular driving [Fig.~\ref{f4}(d)], Floquet hybridization alters the low-energy bands through virtual photon processes, but moving the band minimum past $\mu$ demands either large field amplitudes near the high-frequency limit (see SM~\cite{SM}) or adjustment of $\mu$. Both approaches are intrinsically fragile: the chemical potential is sensitive to disorder and thermal broadening, while single-color Floquet band shifts require amplitudes approaching the limits of the high-frequency regime. Consequently, on–off switching of the RKKY interaction via $\mu$ tuning or large $\mathcal{A}$ is highly parameter-sensitive and difficult to control systematically. For completeness, Fig.~\ref{f7} illustrates how weak doping influences the switching behavior under both single- and two-color driving. In contrast, bichromatic driving [Figs.~\ref{f4}(e)--(h)] offers an intrinsic and efficient control mechanism: interference between photon sectors coherently shifts the global minimum of the lower Floquet band upward, enabling a controlled Lifshitz-like disappearance of the Fermi surface with just two weak beams. The Lifshitz-like transition occurs when $\min_{\mathbf{k}} E_-(\mathbf{k},\mathcal{S}_{\rm c}) = \mu$. Let $\mathbf{k}_0(\mathcal{S})$ denote the momentum minimizing $E_-(\mathbf{k})$, $\nabla_{\mathbf{k}} E_-(\mathbf{k},\mathcal{S})|_{\mathbf{k}=\mathbf{k}_0}=0$. Then $\mathcal{S}_{\rm c}$ satisfies
			\begin{equation} \label{eq:ScImplicit}
				\varepsilon k_0^2 + \mu_\omega(\mathcal{S}_{\rm c}) = \mu + \sqrt{|\boldsymbol{h}(\mathbf{k}_0,\mathcal{S}_{\rm c})|^2}.
			\end{equation}
			For $\lambda_{\rm R}=0.3$ eV$\cdot$\AA, $M_{\rm d}=0.7$, $\theta_M^{\rm d}=0$, $\mu=0$, $\eta_1=\eta_2=+1$, $\hbar\omega=3$ eV, and $\mathcal{A}=0.15$, we find $S_{\rm c}\simeq 2$ for the $d_{x^2-y^2}$-wave altermagnet, consistent with Fig.~\ref{f2} and SM videos~\cite{SM}. Thus, this Lifshitz-like disappearance serves as the main mechanism driving the on–off switching of the RKKY interaction. 
			
			Importantly, this mechanism is robust across different altermagnetic symmetries. Although the lower band exhibits distinct energy distributions across the Brillouin zone for $d_{x^2-y^2}$- and $d_{xy}$-wave altermagnets, see Figs.~\ref{f4}(i)--(l), the critical value $S_{\rm c}$ at which the bands are shifted entirely above $\mu$ is nearly the same in both cases. Thus, for energies below $\mu$, one expects different oscillatory behaviors of the RKKY interaction, whereas the on--off switching process is essentially identical for the two symmetries. These findings are expected to generalize to other $g$-, $f$-, and $i$-wave forms~\cite{tm58-lbdl}. The required amplitude of the second laser beam to achieve an upward shift of the lower band may vary with symmetry, but the essential mechanism—a robust and controllable on–off modulation of the RKKY interaction using bichromatic laser driving—remains valid.\begin{figure}[t]
				\centering
				\includegraphics[width=0.95\linewidth]{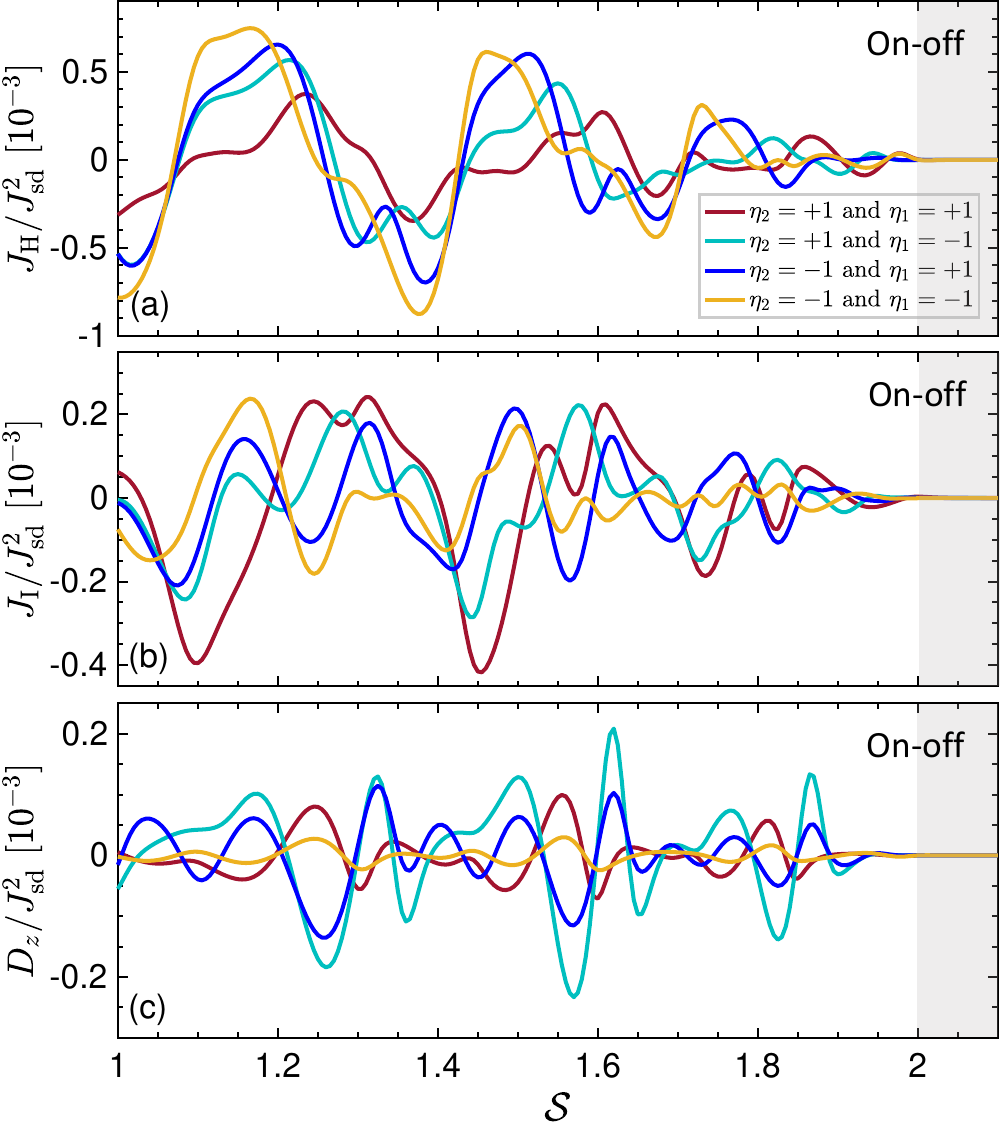}
				\caption{Dependence of the RKKY interaction components, (a) Heisenberg $J_{\rm H}$, (b) Ising $J_{\rm I}$, and (c) DM $D_z$ on the relative amplitude
					$\mathcal{S}$ of the second laser beam for fixed impurity separation
					$R=35$ \AA\, for equal (opposite) chiralities of the two beams. All configurations exhibit nearly the same critical value of $\mathcal{S}_{\rm c} \simeq 2$ at which the RKKY interaction rapidly vanishes. Parameters are $\lambda_{\mathrm{R}}=0.3~\mathrm{eV\cdot}$\AA,
					$M_{\mathrm{d}}=0.7$, $\theta_M^{\mathrm{d}}=0$, and $\mathcal{A}=0.15$.
				}
				\label{f5}
			\end{figure}
			
			We further note that $\mathcal{S}_{\rm c}$ is robust against weak nonmagnetic disorder that preserves the average crystal symmetry and does not qualitatively alter the band structure. Moderate disorder broadens quasiparticle states but leaves the band minimum unchanged, so the Lifshitz-like disappearance of the Fermi surface remains well defined as long as the broadening is smaller than the Floquet gap and spin--orbit energy scales. Finite temperature also smooths the Fermi edge, reducing crossover sharpness but not eliminating it, provided $k_{\rm B}T$ is below the energy separation between the band minimum and the chemical potential near $\mathcal{S}_{\rm c}$. In this regime, Fermi-surface contributions to the RKKY interaction remain suppressed, while residual exchange via virtual Floquet processes persists. As a result, $\mathcal{S}_{\rm c}$ represents a robust crossover scale rather than a fine-tuned critical point, and the on--off switching mechanism survives under realistic disorder and finite-temperature conditions.
			
			Since $D_x$ and $D_y$ appear in both driving schemes, while $D_z$ is unique to the two-color case, we take $D_z$ as the representative DM component in the following.
			
			Figure~\ref{f5} shows how bichromatic driving tunes the strength and magnetic alignment of the impurities via the relative beam amplitude $\mathcal{S}$ at fixed impurity separation $R=35$~\AA, focusing on various chiralities between the beams. For most $\mathcal{S}$, all exchange channels remain finite, indicating coherent spin mediation at the Fermi level. As $\mathcal{S}$ increases, the interactions first show nonmonotonic oscillations, then rapidly suppress at a critical ratio $\mathcal{S}_{\rm c} \simeq 2$ for both equal and opposite chiralities. It is clear that the chirality of the two beams can switch the interaction between FM and AFM in $J_{\rm H}$ and $J_{\rm I}$, as well as toggle the magnetic phase of the two impurities between CW and CCW in $D_z$. The strong dependence on laser chirality reflects the symmetry-selective nature of bichromatic Floquet engineering, providing an additional control over the magnitude and sign of the RKKY interaction.\begin{figure}[t]
				\centering
				\includegraphics[width=0.95\linewidth]{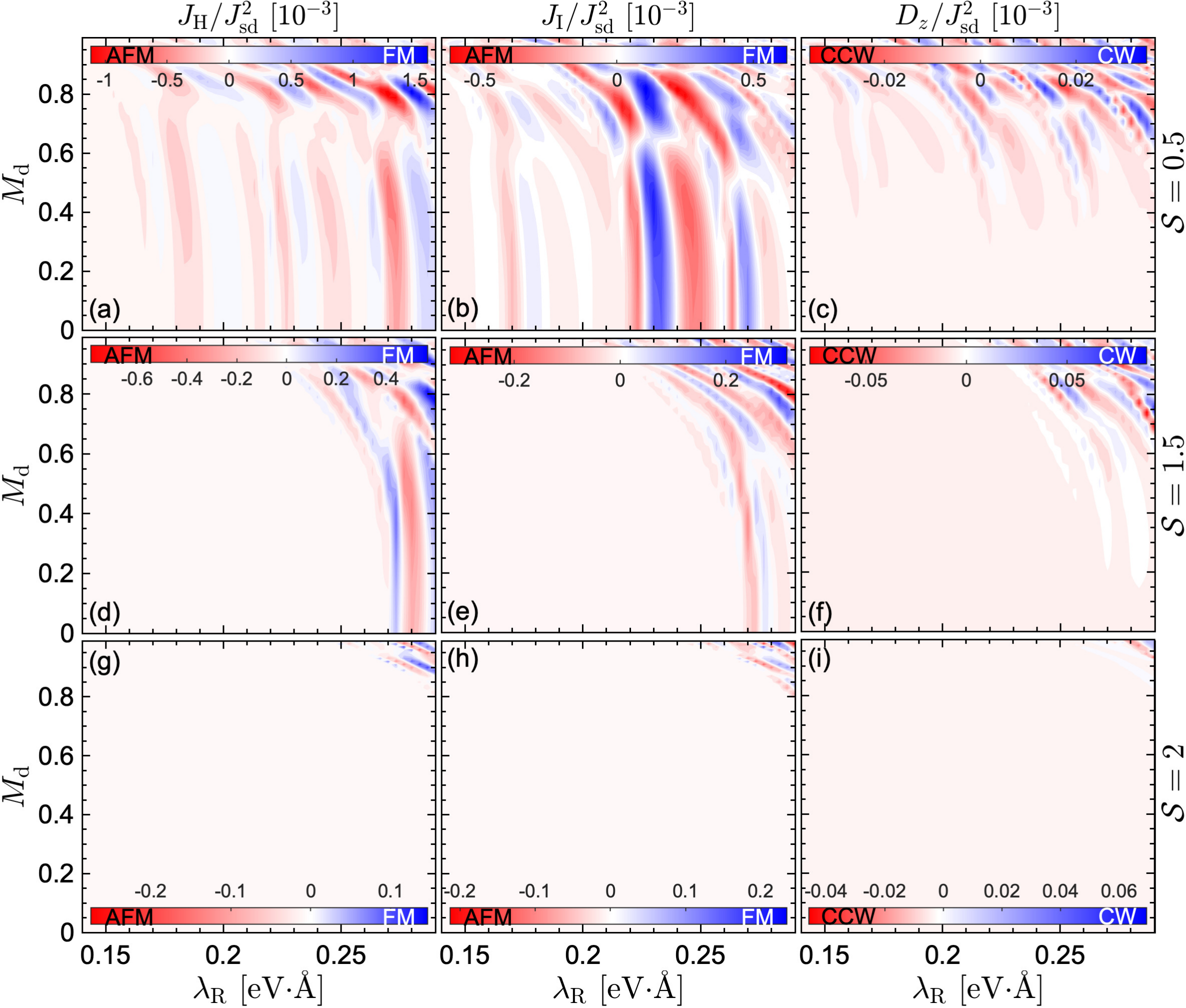}
				\caption{RKKY interaction components as functions of Rashba coupling $\lambda_{\rm R}$ and altermagnetic strength $M_{\rm d}$ for $R=35~\text{\AA}$ and primary light amplitude $\mathcal{A}=0.15$. Columns (left to right) show Heisenberg $J_{\rm H}$, Ising $J_{\rm I}$, and out-of-plane DM $D_z$. Rows (top to bottom) correspond to two-color ratios $\mathcal{S}=0.5$, $1.5$, and $2$. Red (blue) indicates AFM (FM) alignment for $J_{\rm H}$ and $J_{\rm I}$, and CCW (CW) chirality for $D_z$. Increasing $\mathcal{S}$ progressively suppresses all components, nearly switching off the RKKY interaction, except for narrow surviving regions at $\lambda_{\rm R} \gtrsim 0.25$~eV$\cdot$\AA\, in a strong altermagnet.
				}
				\label{f6}
			\end{figure} 
			
			Next, Fig.~\ref{f6} maps the phase diagram $(M_{\rm d},\lambda_{\rm R})$ for three two-color amplitude ratios $\mathcal{S}$, delineating regimes with or without available Fermi-level states and the corresponding suppression of the RKKY interaction, together with the resulting FM/AFM and CW/CCW impurity alignments. For weak driving ($\mathcal{S}=0.5$), all components oscillate, reflecting Fermi-level states mediating long-range exchange. Alternating FM/AFM and CW/CCW regions arise from interference between Rashba spin--momentum locking and the altermagnetic field, while sizable $D_z$ appears only under strong altermagnetic order and RSOC in the two-color driving scheme due to broken effective symmetries. At intermediate driving ($\mathcal{S}=1.5$), the regions with finite exchange shrink, with oscillations confined to narrow bands at larger $\lambda_{\rm R}$ and $M_{\rm d}$, reflecting the Floquet-induced upward shift of the band minimum and depletion of low-energy states. Suppression of all components indicates a common electronic origin tied to Fermi-surface availability. For strong driving ($\mathcal{S}=2$), the RKKY interaction is nearly extinguished, with residual signals only at extreme parameter edges, confirming robust Floquet-induced switching. The concurrent disappearance of FM/AFM and CW/CCW regions confirms suppression arises from the genuine loss of Fermi-level states rather than channel cancellation.
			
			So far, all results were obtained at charge neutrality ($\mu=0$). We now examine the effects of dilute hole ($\mu<0$) and electron ($\mu>0$) doping, focusing on the representative Heisenberg interaction $J_{\rm H}$ in Fig.~\ref{f7}. Figure~\ref{f7} summarizes how the chemical potential $\mu$ controls Floquet-driven tuning of the RKKY interaction under single- and two-color driving. At charge neutrality, $J_{\rm H}$ exhibits oscillations followed by rapid suppression beyond a critical threshold. Oscillations are more regular under $\mathcal A$-driving, while they are sharper under $\mathcal S$-driving, highlighting the efficiency of bichromatic interference in reshaping Floquet bands.  
			
			Hole doping ($\mu<0$) accelerates the extinction of $J_{\rm H}$, which collapses at much smaller $\mathcal A$ and $\mathcal S$ due to an upward shift of Floquet bands relative to the Fermi level. Electron doping ($\mu>0$) enhances $J_{\rm H}$ and extends its range, as Floquet-renormalized conduction-band states continue to mediate long-range exchange.  Notably, for $\mu>0$, single-color driving requires a strong amplitude ($\mathcal A \gtrsim 0.5$) to suppress $J_{\rm H}$, whereas two-color driving achieves on--off switching at much weaker fields. For example, with $\mathcal A=0.15$, tuning $\mathcal S$ to $\mathcal S_{\rm c} \simeq 2.6$ suffices to shift the lower Floquet bands above the chemical potential and suppress the exchange. This efficiency arises from interference between the two frequencies, which selectively redistributes spectral weight and opens low-energy gaps without large overall amplitudes. Therefore, even within the same doping protocol, two-color laser driving provides a substantially more efficient and experimentally favorable route to control and switch the RKKY interaction.
			
			\section{Experimental relevance and material platforms}\label{s5}
			Our results are directly relevant to a growing class of 2D systems combining RSOC with altermagnetic order. Promising platforms include altermagnetic thin films and surfaces derived from recently identified bulk altermagnets, as well as engineered heterostructures where altermagnetic layers are interfaced with heavy-element substrates to enhance RSOC. Candidate materials include KV$_2$Se$_2$O~\cite{Jiang2025}, RbV$_2$Te$_2$O~\cite{Zhang2025_2}, $\kappa$-CL~\cite{Naka2019}, and RuO$_2$~\cite{Feng2022,doi:10.1126/sciadv.aaz8809,PhysRevLett.128.197202,weber2024opticalexcitationspinpolarization,doi:10.1126/sciadv.adj4883}, though the latter remains debated.  The RKKY interaction can be realized experimentally using localized magnetic moments introduced as adatoms or substitutional impurities on the host surface. Such spins can be individually positioned and probed via scanning tunneling microscopy (STM)~\cite{PhysRevLett.98.056601}, allowing direct access to the spatial structure and strength of the indirect exchange. 
			
			Floquet control and on--off switching of the RKKY interaction can be detected using several complementary probes. In STM setups, laser-induced changes would manifest as modifications of impurity-induced bound states or spin-excitation spectra. At higher impurity densities, switching could appear indirectly through altered magnetic ordering or correlation lengths. Time-resolved pump--probe techniques further allow direct tracking of nonequilibrium exchange dynamics via transient spin responses following bichromatic excitation~\cite{Andiel_1999,Cormier2000,PhysRevLett.94.243901,PhysRevA.73.063823,PhysRevLett.110.233903,Kroh18,Jin15,Jin2014,Jin20014,RAJPOOT2024129241}. A concrete realization would then involve $\omega$--$2\omega$ bicircular pumping combined with STM-based impurity spectroscopy on altermagnetic thin films.
			
			\section{Summary and outlook}\label{s6}
            In real experimental setups, magnetic impurities can mask the intrinsic spin-dependent properties of the host material. Moreover, because these impurities interact indirectly through the RKKY mechanism, the performance of applications based on single-spin impurities is inherently limited. This highlights the need for an efficient control knob that can both reveal the pristine electronic and spin features of the host and enable practical single-impurity applications. Such a control scheme should be capable of suppressing the RKKY interaction as much as possible, allowing the intrinsic host information to be extracted while effectively decoupling impurity spins for qubit, sensing, or spintronic applications.
            
			In this work, we have shown that bichromatic laser driving enables efficient, symmetry-selective control of the RKKY interaction in Rashba altermagnets. Our analysis is based on a high-frequency Floquet expansion and assumes weak impurity–electron coupling, such that the RKKY interaction is well described by second-order perturbation theory. By deriving the effective Floquet Hamiltonian and computing real-space Green’s functions, we showed that the interplay between altermagnetic order and light-induced spin–orbit renormalization drives a controllable, Lifshitz-like disappearance of the Fermi surface, allowing robust on--off switching of indirect magnetic exchange at substantially reduced peak field strengths compared to monochromatic driving.  The observed transition differs from an equilibrium Lifshitz transition driven by changes in band topology or carrier density. Instead, it corresponds to a Floquet-induced band-edge crossing controlled dynamically by external fields, without altering the underlying lattice symmetry or requiring fine-tuning of the chemical potential.
            
            This mechanism is insensitive to impurity orientation and altermagnetic symmetry, and naturally generates exchange channels, such as an out-of-plane DM interaction, that are forbidden in equilibrium or under single-color illumination.  Notably, the resulting RKKY interaction is also insensitive to the relative phase of the two driving fields, reflecting a cancellation of phase-dependent contributions arising from the combined role of Green’s functions at $\mathbf{R}$ and $-\mathbf{R}$ in the exchange integrals.
            
            Our analysis further establishes altermagnetism as a key ingredient for realizing nontrivial and controllable RKKY textures under two-color driving. In the absence of altermagnetic order, the Green’s function remains diagonal in spin space, leading only to conventional, collinear exchange interactions similar to those in massive 2D Rashba electron gases. By contrast, finite altermagnetic order generates in-plane spin–orbit fields and anisotropic exchange channels, which are directly responsible for the enhanced tunability and robustness of the laser-controlled switching.  Our results establish bichromatic Floquet engineering as a versatile route for dynamically controlling long-range magnetic interactions in low-dimensional quantum materials.
			
			The results presented here open several avenues for future theoretical and experimental research. The bichromatic Floquet framework can be extended to other magnetic platforms where indirect exchange is important, including antiferromagnets, ferrimagnets, and noncollinear metals. In particular, applying two-color driving to systems with Dirac or multi-valley band structures could enable valley- or sublattice-selective RKKY switching, providing additional control beyond altermagnets. While moderate disorder and finite temperature are not expected to qualitatively alter the Floquet-induced band shifts, strong scattering or low-frequency driving may invalidate the effective Hamiltonian description. Exploring these regimes remains an interesting direction for future work. 
			
			\section*{Acknowledgments}
			M. Y. and J. K. F. were supported by the Department of Energy, Office of Basic Energy Sciences, Division of Materials Sciences and Engineering under Contract No. DE-FG02-08ER46542 for the formal developments, the numerical work, and the writing of the manuscript. J. F. K. was also supported by the McDevitt bequest at Georgetown University. P.-H. F. acknowledges the financial support from the Fondazione Cariplo under the grant 2023-2594 for preliminary analytical calculations and discussion. M. Y. and P.-H. F. thank S. A. A. Ghorashi for useful discussions.
			
			\section*{Data availability}
			The data supporting the findings of this article are openly available in the Zenodo database~\cite{Zenodo}.

			\appendix
			
			\section{\rd{Robustness of Floquet switching against dissipation}}\label{ap1}
			\rd{In a realistic material setting, coupling to a macroscopic fermionic reservoir (e.g., a substrate) induces dissipation, which can lead to nonthermal occupations and finite quasiparticle lifetimes. To rigorously test the robustness of our central on-off switching mechanism against these dissipative effects, we recalculate the RKKY interaction by introducing a phenomenological scattering rate $\Gamma$ into our real-space retarded Floquet Green's function formalism [Eq.~\eqref{eq_21}]. Using a finite $+i\Gamma$, we model the loss of spatial coherence and quasiparticle scattering.}\begin{figure}[t]
				\centering
				\includegraphics[width=0.95\linewidth]{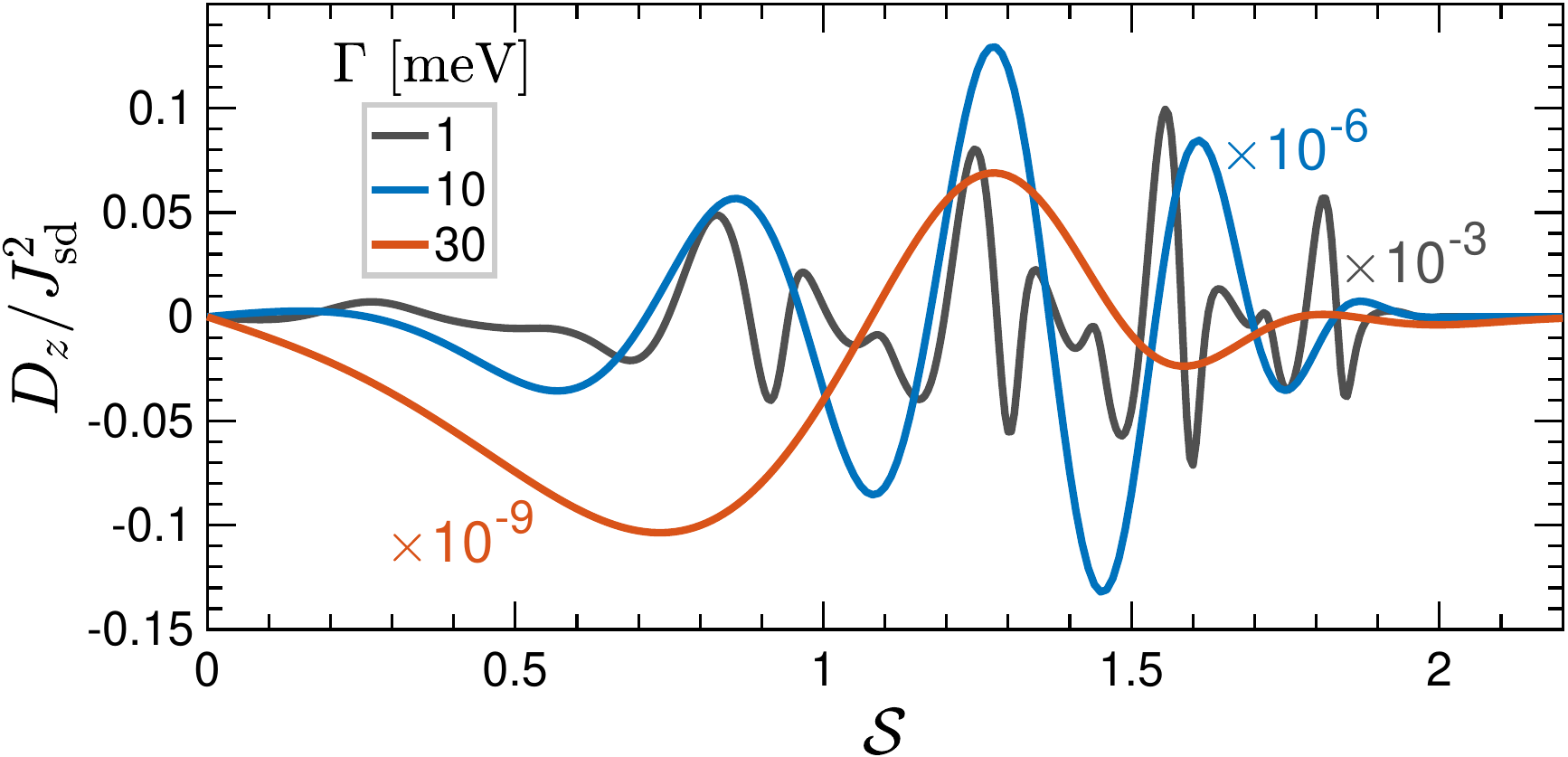}
				\caption{\rd{Dependence of the out-of-plane DM interaction component $D_z$ on the relative amplitude of the second laser beam ($\mathcal{S}$) for varying phenomenological broadening parameters $\Gamma = 1$~meV, $10$~meV, and $30$~meV. To allow visual comparison on the same axis, the curves are scaled by $10^{-3}$, $10^{-6}$, and $10^{-9}$, respectively. While increased dissipation exponentially dampens the absolute magnitude of the exchange coupling and smooths out the high-frequency quantum interference oscillations, the critical on-off switching threshold near $\mathcal{S}_c \simeq 2$ remains robustly well-defined.}}
				\label{fig:dissipation}
			\end{figure}
				
			\rd{The results of this calculation for the representative out-of-plane DM component, $D_z$, are shown in Fig.~\ref{fig:dissipation} for $\Gamma = 1$~meV, $10$~meV, and $30$~meV. As the scattering rate increases, the absolute magnitude of the RKKY interaction drops precipitously. This exponential suppression is a known physical consequence of a finite electronic mean free path; the spatial coherence required to mediate long-range magnetic interactions is heavily damped by the bath. Furthermore, at higher broadening ($\Gamma = 30$~meV), the rapid, high-frequency oscillations as a function of the drive ratio $\mathcal{S}$ are washed out, reflecting the smearing of the Fermi surface and the suppression of delicate multi-photon interference pathways.}
				
			\rd{Most importantly, despite the massive reduction in overall magnitude and the smoothing of the oscillatory structure, the critical on-off switching threshold remains securely pinned near $\mathcal{S}_c \simeq 2$ for all values of $\Gamma$. The proposed switching mechanism relies on the Floquet-induced upward shift of the global band minimum past the chemical potential. Because this macroscopic band shift is driven by energy scales ($\hbar\omega \ge 1$~eV and corresponding laser amplitudes) that are significantly larger than the scattering rate $\Gamma$, the Lifshitz-like disappearance of the Fermi surface survives the thermal and phenomenological smearing. Thus, the fundamental capability of bichromatic driving to switch off the RKKY interaction remains completely robust in a dissipative environment.}
			
			\section{\rd{Verification of the truncated effective Hamiltonian}}\label{ap2}
			\begin{figure}[b]
				\centering
				\includegraphics[width=0.95\linewidth]{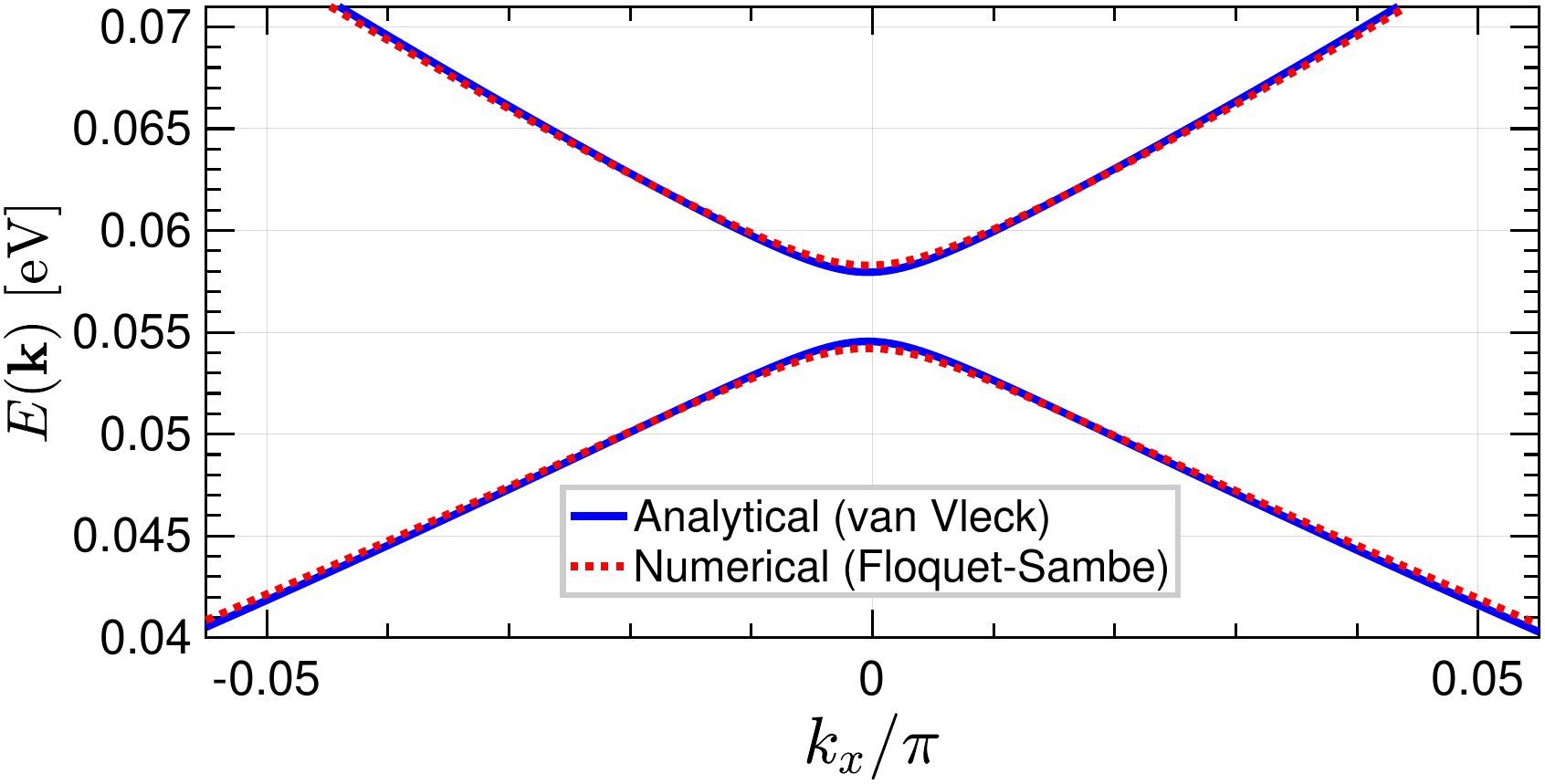}
				\caption{\rd{Comparison of the quasienergy spectrum along the $k_x$ direction (at $k_y = 0$) computed via the analytical $1/\omega$ van Vleck expansion (solid blue line) and the exact numerical Floquet-Sambe diagonalization (dotted red line). The parameters are set near the critical switching threshold $\mathcal{S}_{\rm c} = 2$, with the primary field amplitude $\mathcal{A} = 0.15$ and driving frequency $\hbar\omega = 3$ eV. The near-perfect overlap validates the truncation used to derive the effective Floquet Hamiltonian.}}
				\label{fig:floquet_sambe}
			\end{figure}
			
			\rd{Under the bichromatic driving protocol proposed in this work, the suppression of the RKKY interaction occurs at a relatively large relative amplitude of the second-harmonic beam, $\mathcal{S}_c \simeq 2$. To ensure that the analytical truncation used in the high-frequency van Vleck expansion in Eq.~\eqref{eq_4} remains valid in this regime, and to verify that all relevant Fourier components are properly included, we compare our effective Hamiltonian against a full numerical Floquet-Sambe calculation. The numerical Floquet-Sambe approach~\cite{keliri2026sambeapproachfloquetlindbladopen} constructs the exact block-matrix representation of the time-periodic Hamiltonian in the extended energy-photon Hilbert space. By diagonalizing this infinite-dimensional matrix—truncated at a sufficiently large photon number cutoff to ensure strict convergence—we obtain the exact quasienergy spectrum without relying on a perturbative $1/\omega$ expansion. Figure~\ref{fig:floquet_sambe} shows the quasienergy band structure along the $k_x$ direction (at $k_y = 0$) calculated near the critical switching threshold $\mathcal{S}_c = 2$. The analytical results (solid blue line) map almost perfectly onto the exact numerical diagonalization (dotted red line). }
			
			\rd{This excellent quantitative agreement confirms several key physical points: (i) The macroscopic upward shift of the global band minimum, which drives the Lifshitz-like disappearance of the Fermi surface and turns off the RKKY interaction, is highly accurate and not an artifact of the truncation. (ii) The dynamically generated in-plane Zeeman-like fields ($\Omega_x, \Omega_y$) and the out-of-plane DM interaction ($D_z$) are genuine physical features captured perfectly by the leading-order commutators of the drive harmonics. And (iii) because the fundamental driving frequency ($\hbar\omega \ge 1$ eV) heavily outweighs the intrinsic energy scales of the 2D Rashba altermagnet, higher-order multi-photon transitions remain highly off-resonant. Consequently, the $1/\omega$ van Vleck effective Hamiltonian provides a rigorously justified framework for modeling the bichromatic optical control of the RKKY interaction.}
}    

\bibliography{bib.bib}

\end{document}